
%

\input epsf.tex
%
%
%
%

\catcode `\@=11 

\def\@version{1.4}
\def\@verdate{22nd Feb 1994}

%
%
%
%


\newif\ifprod@font

\ifx\@typeface\undefined
  \def\@typeface{Comp. Modern}\prod@fontfalse
\else
  \prod@fonttrue 
\fi

\def\newfam{\alloc@8\fam\chardef\sixt@@n} 

\ifprod@font
\font\fiverm=mtr10 at 5pt
\font\fivebf=mtbx10 at 5pt
\font\fiveit=mtti10 at 5pt
\font\fivesl=mtsl10 at 5pt
\font\fivett=mttt10 at 5pt     \hyphenchar\fivett=-1
\font\fivecsc=mtcsc10 at 5pt
\font\fivesf=mtss10 at 5pt
\font\fivei=mtmi10 at 5pt      \skewchar\fivei='177
\font\fivemib=mtmib10 at 5pt   \skewchar\fivemib='177
\font\fivesy=mtsy10 at 5pt     \skewchar\fivesy='60
\font\fivesyb=mtbsy10 at 5pt   \skewchar\fivesyb='60

\font\sixrm=mtr10 at 6pt
\font\sixbf=mtbx10 at 6pt
\font\sixit=mtti10 at 6pt
\font\sixsl=mtsl10 at 6pt
\font\sixtt=mttt10 at 6pt      \hyphenchar\sixtt=-1
\font\sixcsc=mtcsc10 at 6pt
\font\sixsf=mtss10 at 6pt
\font\sixi=mtmi10 at 6pt       \skewchar\sixi='177
\font\sixmib=mtmib10 at 6pt    \skewchar\sixmib='177
\font\sixsy=mtsy10 at 6pt      \skewchar\sixsy='60
\font\sixsyb=mtbsy10 at 6pt    \skewchar\sixsyb='60

\font\sevenrm=mtr10 at 7pt
\font\sevenbf=mtbx10 at 7pt
\font\sevenit=mtti10 at 7pt
\font\sevensl=mtsl10 at 7pt
\font\seventt=mttt10 at 7pt     \hyphenchar\seventt=-1
\font\sevencsc=mtcsc10 at 7pt
\font\sevensf=mtss10 at 7pt
\font\seveni=mtmi10 at 7pt      \skewchar\seveni='177
\font\sevenmib=mtmib10 at 7pt   \skewchar\sevenmib='177
\font\sevensy=mtsy10 at 7pt     \skewchar\sevensy='60
\font\sevensyb=mtbsy10 at 7pt   \skewchar\sevensyb='60

\font\eightrm=mtr10 at 8pt
\font\eightbf=mtbx10 at 8pt
\font\eightit=mtti10 at 8pt
\font\eighti=mtmi10 at 8pt      \skewchar\eighti='177
\font\eightmib=mtmib10 at 8pt   \skewchar\eightmib='177
\font\eightsy=mtsy10 at 8pt     \skewchar\eightsy='60
\font\eightsyb=mtbsy10 at 8pt   \skewchar\eightsyb='60
\font\eightsl=mtsl10 at 8pt
\font\eighttt=mttt10 at 8pt     \hyphenchar\eighttt=-1
\font\eightcsc=mtcsc10 at 8pt
\font\eightsf=mtss10 at 8pt

\font\ninerm=mtr10 at 9pt
\font\ninebf=mtbx10 at 9pt
\font\nineit=mtti10 at 9pt
\font\ninei=mtmi10 at 9pt      \skewchar\ninei='177
\font\ninemib=mtmib10 at 9pt   \skewchar\ninemib='177
\font\ninesy=mtsy10 at 9pt     \skewchar\ninesy='60
\font\ninesyb=mtbsy10 at 9pt   \skewchar\ninesyb='60
\font\ninesl=mtsl10 at 9pt
\font\ninett=mttt10 at 9pt     \hyphenchar\ninett=-1
\font\ninecsc=mtcsc10 at 9pt
\font\ninesf=mtss10 at 9pt

\font\tenrm=mtr10
\font\tenbf=mtbx10
\font\tenit=mtti10
\font\teni=mtmi10		\skewchar\teni='177
\font\tenmib=mtmib10	\skewchar\tenmib='177
\font\tensy=mtsy10		\skewchar\tensy='60
\font\tensyb=mtbsy10	\skewchar\tensyb='60
\font\tenex=cmex10
\font\tensl=mtsl10
\font\tentt=mttt10		\hyphenchar\tentt=-1
\font\tencsc=mtcsc10
\font\tensf=mtss10

\font\elevenrm=mtr10 at 11pt
\font\elevenbf=mtbx10 at 11pt
\font\elevenit=mtti10 at 11pt
\font\eleveni=mtmi10 at 11pt      \skewchar\eleveni='177
\font\elevenmib=mtmib10 at 11pt   \skewchar\elevenmib='177
\font\elevensy=mtsy10 at 11pt     \skewchar\elevensy='60
\font\elevensyb=mtbsy10 at 11pt   \skewchar\elevensyb='60
\font\elevensl=mtsl10 at 11pt
\font\eleventt=mttt10 at 11pt     \hyphenchar\eleventt=-1
\font\elevencsc=mtcsc10 at 11pt
\font\elevensf=mtss10 at 11pt

\font\twelverm=mtr10 at 12pt
\font\twelvebf=mtbx10 at 12pt
\font\twelveit=mtti10 at 12pt
\font\twelvesl=mtsl10 at 12pt
\font\twelvett=mttt10 at 12pt     \hyphenchar\twelvett=-1
\font\twelvecsc=mtcsc10 at 12pt
\font\twelvesf=mtss10 at 12pt
\font\twelvei=mtmi10 at 12pt      \skewchar\twelvei='177
\font\twelvemib=mtmib10 at 12pt   \skewchar\twelvemib='177
\font\twelvesy=mtsy10 at 12pt     \skewchar\twelvesy='60
\font\twelvesyb=mtbsy10 at 12pt   \skewchar\twelvesyb='60

\font\fourteenrm=mtr10 at 14pt
\font\fourteenbf=mtbx10 at 14pt
\font\fourteenit=mtti10 at 14pt
\font\fourteeni=mtmi10 at 14pt      \skewchar\fourteeni='177
\font\fourteenmib=mtmib10 at 14pt   \skewchar\fourteenmib='177
\font\fourteensy=mtsy10 at 14pt     \skewchar\fourteensy='60
\font\fourteensyb=mtbsy10 at 14pt   \skewchar\fourteensyb='60
\font\fourteensl=mtsl10 at 14pt
\font\fourteentt=mttt10 at 14pt     \hyphenchar\fourteentt=-1
\font\fourteencsc=mtcsc10 at 14pt
\font\fourteensf=mtss10 at 14pt

\font\seventeenrm=mtr10 at 17pt
\font\seventeenbf=mtbx10 at 17pt
\font\seventeenit=mtti10 at 17pt
\font\seventeeni=mtmi10 at 17pt      \skewchar\seventeeni='177
\font\seventeenmib=mtmib10 at 17pt   \skewchar\seventeenmib='177
\font\seventeensy=mtsy10 at 17pt     \skewchar\seventeensy='60
\font\seventeensyb=mtbsy10 at 17pt   \skewchar\seventeensyb='60
\font\seventeensl=mtsl10 at 17pt
\font\seventeentt=mttt10 at 17pt     \hyphenchar\seventeentt=-1
\font\seventeencsc=mtcsc10 at 17pt
\font\seventeensf=mtss10 at 17pt


\newfam\xmfam
\newfam\ymfam

\font\fivexm=mtxm10 at 5pt
\font\sixxm=mtxm10 at 6pt
\font\sevenxm=mtxm10 at 7pt
\font\eightxm=mtxm10 at 8pt
\font\ninexm=mtxm10 at 9pt
\font\tenxm=mtxm10
\font\elevenxm=mtxm10 at 11pt
\font\twelvexm=mtxm10 at 12pt
\font\fourteenxm=mtxm10 at 14pt
\font\seventeenxm=mtxm10 at 17pt

\font\fiveym=mtym10 at 5pt
\font\sixym=mtym10 at 6pt
\font\sevenym=mtym10 at 7pt
\font\eightym=mtym10 at 8pt
\font\nineym=mtym10 at 9pt
\font\tenym=mtym10
\font\elevenym=mtym10 at 11pt
\font\twelveym=mtym10 at 12pt
\font\fourteenym=mtym10 at 14pt
\font\seventeenym=mtym10 at 17pt
\else
\font\fiverm=cmr5
\font\fivei=cmmi5             \skewchar\fivei='177
\font\fivemib=cmmib10 at 5pt  \skewchar\fivemib='177
\font\fivesy=cmsy5            \skewchar\fivesy='60
\font\fivesyb=cmbsy10 at 5pt  \skewchar\fivesyb='60
\font\fivebf=cmbx5

\font\sixrm=cmr6
\font\sixi=cmmi6             \skewchar\sixi='177
\font\sixmib=cmmib10 at 6pt  \skewchar\sixmib='177
\font\sixsy=cmsy6            \skewchar\sixsy='60
\font\sixsyb=cmbsy10 at 6pt  \skewchar\sixsyb='60
\font\sixbf=cmbx6

\font\sevenrm=cmr7
\font\seveni=cmmi7             \skewchar\seveni='177
\font\sevenmib=cmmib10 at 7pt  \skewchar\sevenmib='177
\font\sevensy=cmsy7            \skewchar\sevensy='60
\font\sevensyb=cmbsy10 at 7pt  \skewchar\sevensyb='60
\font\sevenbf=cmbx7

\font\eightrm=cmr8
\font\eightbf=cmbx8
\font\eightit=cmti8
\font\eighti=cmmi8			\skewchar\eighti='177
\font\eightmib=cmmib10 at 8pt	\skewchar\eightmib='177
\font\eightsy=cmsy8			\skewchar\eightsy='60
\font\eightsyb=cmbsy10 at 8pt	\skewchar\eightsyb='60
\font\eightsl=cmsl8
\font\eighttt=cmtt8			\hyphenchar\eighttt=-1
\font\eightcsc=cmcsc10 at 8pt
\font\eightsf=cmss8

\font\ninerm=cmr9
\font\ninebf=cmbx9
\font\nineit=cmti9
\font\ninei=cmmi9			\skewchar\ninei='177
\font\ninemib=cmmib10 at 9pt	\skewchar\ninemib='177
\font\ninesy=cmsy9			\skewchar\ninesy='60
\font\ninesyb=cmbsy10 at 9pt	\skewchar\ninesyb='60
\font\ninesl=cmsl9
\font\ninett=cmtt9			\hyphenchar\ninett=-1
\font\ninecsc=cmcsc10 at 9pt
\font\ninesf=cmss9

\font\tenrm=cmr10
\font\tenbf=cmbx10
\font\tenit=cmti10
\font\teni=cmmi10		\skewchar\teni='177
\font\tenmib=cmmib10	\skewchar\tenmib='177
\font\tensy=cmsy10		\skewchar\tensy='60
\font\tensyb=cmbsy10	\skewchar\tensyb='60
\font\tenex=cmex10
\font\tensl=cmsl10
\font\tentt=cmtt10		\hyphenchar\tentt=-1
\font\tencsc=cmcsc10
\font\tensf=cmss10

\font\elevenrm=cmr10 scaled \magstephalf
\font\elevenbf=cmbx10 scaled \magstephalf
\font\elevenit=cmti10 scaled \magstephalf
\font\eleveni=cmmi10 scaled \magstephalf	\skewchar\eleveni='177
\font\elevenmib=cmmib10 scaled \magstephalf	\skewchar\elevenmib='177
\font\elevensy=cmsy10 scaled \magstephalf	\skewchar\elevensy='60
\font\elevensyb=cmbsy10 scaled \magstephalf	\skewchar\elevensyb='60
\font\elevensl=cmsl10 scaled \magstephalf
\font\eleventt=cmtt10 scaled \magstephalf	\hyphenchar\eleventt=-1
\font\elevencsc=cmcsc10 scaled \magstephalf
\font\elevensf=cmss10 scaled \magstephalf

\font\twelverm=cmr10 scaled \magstep1
\font\twelvebf=cmbx10 scaled \magstep1
\font\twelvei=cmmi10 scaled \magstep1      \skewchar\twelvei='177
\font\twelvemib=cmmib10 scaled \magstep1   \skewchar\twelvemib='177
\font\twelvesy=cmsy10 scaled \magstep1     \skewchar\twelvesy='60
\font\twelvesyb=cmbsy10 scaled \magstep1   \skewchar\twelvesyb='60

\font\fourteenrm=cmr10 scaled \magstep2
\font\fourteenbf=cmbx10 scaled \magstep2
\font\fourteenit=cmti10 scaled \magstep2
\font\fourteeni=cmmi10 scaled \magstep2		\skewchar\fourteeni='177
\font\fourteenmib=cmmib10 scaled \magstep2	\skewchar\fourteenmib='177
\font\fourteensy=cmsy10 scaled \magstep2	\skewchar\fourteensy='60
\font\fourteensyb=cmbsy10 scaled \magstep2	\skewchar\fourteensyb='60
\font\fourteensl=cmsl10 scaled \magstep2
\font\fourteentt=cmtt10 scaled \magstep2	\hyphenchar\fourteentt=-1
\font\fourteencsc=cmcsc10 scaled \magstep2
\font\fourteensf=cmss10 scaled \magstep2

\font\seventeenrm=cmr10 scaled \magstep3
\font\seventeenbf=cmbx10 scaled \magstep3
\font\seventeenit=cmti10 scaled \magstep3
\font\seventeeni=cmmi10 scaled \magstep3	\skewchar\seventeeni='177
\font\seventeenmib=cmmib10 scaled \magstep3	\skewchar\seventeenmib='177
\font\seventeensy=cmsy10 scaled \magstep3	\skewchar\seventeensy='60
\font\seventeensyb=cmbsy10 scaled \magstep3	\skewchar\seventeensyb='60
\font\seventeensl=cmsl10 scaled \magstep3
\font\seventeentt=cmtt10 scaled \magstep3	\hyphenchar\seventeentt=-1
\font\seventeencsc=cmcsc10 scaled \magstep3
\font\seventeensf=cmss10 scaled \magstep3
\fi

\def\hexnumber#1{\ifcase#1 0\or1\or2\or3\or4\or5\or6\or7\or8\or9\or
  A\or B\or C\or D\or E\or F\fi}

\ifprod@font
  \edef\@xm{\hexnumber\xmfam}
  \edef\@ym{\hexnumber\ymfam}
\fi

\def\mib{\hexnumber\mibfam}
\def\syb{\hexnumber\sybfam}

\def\makestrut{%
  \setbox\strutbox=\hbox{%
    \vrule height.7\baselineskip depth.3\baselineskip width \z@}%
}

\def\baselinestretch{1}
\newskip\tmp@bls

\def\b@ls#1{
  \tmp@bls=#1\relax
  \baselineskip=#1\relax\makestrut
  \normalbaselineskip=\baselinestretch\tmp@bls
  \normalbaselines
}

\def\nostb@ls#1{
  \normalbaselineskip=#1\relax
  \normalbaselines
  \makestrut
}

%

\newfam\mibfam 
\newfam\sybfam 
\newfam\scfam  
\newfam\sffam  

\def\mit{\fam\@ne}

\def\cal{\fam\tw@}

\def\em{\ifdim\fontdimen1\font>\z@ \rm\else\it\fi}

\textfont3=\tenex
\scriptfont3=\tenex
\scriptscriptfont3=\tenex

\setbox0=\hbox{\tenex B} \p@renwd=\wd0 

\def\eightpoint{
  \def\rm{\fam0\eightrm}%
  \textfont0=\eightrm \scriptfont0=\sixrm \scriptscriptfont0=\fiverm%
  \textfont1=\eighti  \scriptfont1=\sixi  \scriptscriptfont1=\fivei%
  \textfont2=\eightsy \scriptfont2=\sixsy \scriptscriptfont2=\fivesy%
  \textfont\itfam=\eightit\def\it{\fam\itfam\eightit}%
  \ifprod@font
    \scriptfont\itfam=\sixit
      \scriptscriptfont\itfam=\fiveit
  \else
    \scriptfont\itfam=\eightit
      \scriptscriptfont\itfam=\eightit
  \fi
  \textfont\bffam=\eightbf%
    \scriptfont\bffam=\sixbf%
      \scriptscriptfont\bffam=\fivebf%
  \def\bf{\fam\bffam\eightbf}%
  \textfont\slfam=\eightsl\def\sl{\fam\slfam\eightsl}%
  \ifprod@font
    \scriptfont\slfam=\sixsl
      \scriptscriptfont\slfam=\fivesl
  \else
    \scriptfont\slfam=\eightsl
      \scriptscriptfont\slfam=\eightsl
  \fi
  \textfont\ttfam=\eighttt\def\tt{\fam\ttfam\eighttt}%
  \ifprod@font
    \scriptfont\ttfam=\sixtt
      \scriptscriptfont\ttfam=\fivett
  \else
    \scriptfont\ttfam=\eighttt
      \scriptscriptfont\ttfam=\eighttt
  \fi
  \textfont\scfam=\eightcsc\def\sc{\fam\scfam\eightcsc}%
  \ifprod@font
    \scriptfont\scfam=\sixcsc
      \scriptscriptfont\scfam=\fivecsc
  \else
    \scriptfont\scfam=\eightcsc
      \scriptscriptfont\scfam=\eightcsc
  \fi
  \textfont\sffam=\eightsf\def\sf{\fam\sffam\eightsf}%
  \ifprod@font
    \scriptfont\sffam=\sixsf
      \scriptscriptfont\sffam=\fivesf
  \else
    \scriptfont\sffam=\eightsf
      \scriptscriptfont\sffam=\eightsf
  \fi
  \textfont\mibfam=\eightmib
    \scriptfont\mibfam=\sixmib
      \scriptscriptfont\mibfam=\fivemib
  \textfont\sybfam=\eightsyb
    \scriptfont\sybfam=\sixsyb
      \scriptscriptfont\sybfam=\fivesyb
  \ifprod@font
    \textfont\xmfam=\eightxm
      \scriptfont\xmfam=\sixxm
        \scriptscriptfont\xmfam=\fivexm
    \textfont\ymfam=\eightym
      \scriptfont\ymfam=\sixym
        \scriptscriptfont\ymfam=\fiveym
  \fi
  \def\oldstyle{\fam\@ne\eighti}%
  \def\boldstyle{\fam\mibfam\eightmib}%
  \b@ls{10pt}\rm%
}

\def\ninepoint{
  \def\rm{\fam0\ninerm}%
  \textfont0=\ninerm \scriptfont0=\sixrm \scriptscriptfont0=\fiverm%
  \textfont1=\ninei  \scriptfont1=\sixi  \scriptscriptfont1=\fivei%
  \textfont2=\ninesy \scriptfont2=\sixsy \scriptscriptfont2=\fivesy%
  \textfont\itfam=\nineit\def\it{\fam\itfam\nineit}%
  \ifprod@font
    \scriptfont\itfam=\sixit
      \scriptscriptfont\itfam=\fiveit
  \else
    \scriptfont\itfam=\nineit
      \scriptscriptfont\itfam=\nineit
  \fi
  \textfont\bffam=\ninebf%
    \scriptfont\bffam=\sixbf%
      \scriptscriptfont\bffam=\fivebf%
  \def\bf{\fam\bffam\ninebf}%
  \textfont\slfam=\ninesl\def\sl{\fam\slfam\ninesl}%
  \ifprod@font
    \scriptfont\slfam=\sixsl
      \scriptscriptfont\slfam=\fivesl
  \else
    \scriptfont\slfam=\ninesl
      \scriptscriptfont\slfam=\ninesl
  \fi
  \textfont\ttfam=\ninett\def\tt{\fam\ttfam\ninett}%
  \ifprod@font
    \scriptfont\ttfam=\sixtt
      \scriptscriptfont\ttfam=\fivett
  \else
    \scriptfont\ttfam=\ninett
      \scriptscriptfont\ttfam=\ninett
  \fi
  \textfont\scfam=\ninecsc\def\sc{\fam\scfam\ninecsc}%
  \ifprod@font
    \scriptfont\scfam=\sixcsc
      \scriptscriptfont\scfam=\fivecsc
  \else
    \scriptfont\scfam=\ninecsc
      \scriptscriptfont\scfam=\ninecsc
  \fi
  \textfont\sffam=\ninesf\def\sf{\fam\sffam\ninesf}%
  \ifprod@font
    \scriptfont\sffam=\sixsf
      \scriptscriptfont\sffam=\fivesf
  \else
    \scriptfont\sffam=\ninesf
      \scriptscriptfont\sffam=\ninesf
  \fi
  \textfont\mibfam=\ninemib
    \scriptfont\mibfam=\sixmib
      \scriptscriptfont\mibfam=\fivemib
  \textfont\sybfam=\ninesyb
    \scriptfont\sybfam=\sixsyb
      \scriptscriptfont\sybfam=\fivesyb
  \ifprod@font
    \textfont\xmfam=\ninexm
      \scriptfont\xmfam=\sixxm
        \scriptscriptfont\xmfam=\fivexm
    \textfont\ymfam=\nineym
      \scriptfont\ymfam=\sixym
        \scriptscriptfont\ymfam=\fiveym
  \fi
  \def\oldstyle{\fam\@ne\ninei}%
  \def\boldstyle{\fam\mibfam\ninemib}%
  \b@ls{\TextLeading plus \Feathering}\rm%
}

\def\tenpoint{
  \def\rm{\fam0\tenrm}%
  \textfont0=\tenrm \scriptfont0=\sevenrm \scriptscriptfont0=\fiverm%
  \textfont1=\teni  \scriptfont1=\seveni  \scriptscriptfont1=\fivei%
  \textfont2=\tensy \scriptfont2=\sevensy \scriptscriptfont2=\fivesy%
  \textfont\itfam=\tenit\def\it{\fam\itfam\tenit}%
  \ifprod@font
    \scriptfont\itfam=\sevenit
      \scriptscriptfont\itfam=\fiveit
  \else
    \scriptfont\itfam=\tenit
      \scriptscriptfont\itfam=\tenit
  \fi
  \textfont\bffam=\tenbf%
    \scriptfont\bffam=\sevenbf%
      \scriptscriptfont\bffam=\fivebf%
  \def\bf{\fam\bffam\tenbf}%
  \textfont\slfam=\tensl\def\sl{\fam\slfam\tensl}%
  \ifprod@font
    \scriptfont\slfam=\sevensl
      \scriptscriptfont\slfam=\fivesl
  \else
    \scriptfont\slfam=\tensl
      \scriptscriptfont\slfam=\tensl
  \fi
  \textfont\ttfam=\tentt\def\tt{\fam\ttfam\tentt}%
  \ifprod@font
    \scriptfont\ttfam=\seventt
      \scriptscriptfont\ttfam=\fivett
  \else
    \scriptfont\ttfam=\tentt
      \scriptscriptfont\ttfam=\tentt
  \fi
  \textfont\scfam=\tencsc\def\sc{\fam\scfam\tencsc}%
  \ifprod@font
    \scriptfont\scfam=\sevencsc
      \scriptscriptfont\scfam=\fivecsc
  \else
    \scriptfont\scfam=\tencsc
      \scriptscriptfont\scfam=\tencsc
  \fi
  \textfont\sffam=\tensf\def\sf{\fam\sffam\tensf}%
  \ifprod@font
    \scriptfont\sffam=\sevensf
      \scriptscriptfont\sffam=\fivesf
  \else
    \scriptfont\sffam=\tensf
      \scriptscriptfont\sffam=\tensf
  \fi
  \textfont\mibfam=\tenmib
    \scriptfont\mibfam=\sevenmib
      \scriptscriptfont\mibfam=\fivemib
  \textfont\sybfam=\tensyb
    \scriptfont\sybfam=\sevensyb
      \scriptscriptfont\sybfam=\fivesyb
  \ifprod@font
    \textfont\xmfam=\tenxm
      \scriptfont\xmfam=\sevenxm
        \scriptscriptfont\xmfam=\fivexm
    \textfont\ymfam=\tenym
      \scriptfont\ymfam=\sevenym
        \scriptscriptfont\ymfam=\fiveym
  \fi
  \def\oldstyle{\fam\@ne\teni}%
  \def\boldstyle{\fam\mibfam\tenmib}%
  \b@ls{11pt}\rm%
}

\def\elevenpoint{
  \def\rm{\fam0\elevenrm}%
  \textfont0=\elevenrm \scriptfont0=\eightrm \scriptscriptfont0=\sixrm%
  \textfont1=\eleveni  \scriptfont1=\eighti  \scriptscriptfont1=\sixi%
  \textfont2=\elevensy \scriptfont2=\eightsy \scriptscriptfont2=\sixsy%
  \textfont\itfam=\elevenit\def\it{\fam\itfam\elevenit}%
  \ifprod@font
    \scriptfont\itfam=\eightit
      \scriptscriptfont\itfam=\sixit
  \else
    \scriptfont\itfam=\elevenit
      \scriptscriptfont\itfam=\elevenit
  \fi
  \textfont\bffam=\elevenbf%
    \scriptfont\bffam=\eightbf%
      \scriptscriptfont\bffam=\sixbf%
  \def\bf{\fam\bffam\elevenbf}%
  \textfont\slfam=\elevensl\def\sl{\fam\slfam\elevensl}%
  \ifprod@font
    \scriptfont\slfam=\eightsl
      \scriptscriptfont\slfam=\sixsl
  \else
    \scriptfont\slfam=\elevensl
      \scriptscriptfont\slfam=\elevensl
  \fi
  \textfont\ttfam=\eleventt\def\tt{\fam\ttfam\eleventt}%
  \ifprod@font
    \scriptfont\ttfam=\eighttt
      \scriptscriptfont\ttfam=\sixtt
  \else
    \scriptfont\ttfam=\eleventt
      \scriptscriptfont\ttfam=\eleventt
  \fi
  \textfont\scfam=\elevencsc\def\sc{\fam\scfam\elevencsc}%
  \ifprod@font
    \scriptfont\scfam=\eightcsc
      \scriptscriptfont\scfam=\sixcsc
  \else
    \scriptfont\scfam=\elevencsc
      \scriptscriptfont\scfam=\elevencsc
  \fi
  \textfont\sffam=\elevensf\def\sf{\fam\sffam\elevensf}%
  \ifprod@font
    \scriptfont\sffam=\eightsf
      \scriptscriptfont\sffam=\sixsf
  \else
    \scriptfont\sffam=\elevensf
      \scriptscriptfont\sffam=\elevensf
  \fi
  \textfont\mibfam=\elevenmib
    \scriptfont\mibfam=\eightmib
      \scriptscriptfont\mibfam=\sixmib
  \textfont\sybfam=\elevensyb
    \scriptfont\sybfam=\eightsyb
      \scriptscriptfont\sybfam=\sixsyb
  \ifprod@font
    \textfont\xmfam=\elevenxm
      \scriptfont\xmfam=\eightxm
       \scriptscriptfont\xmfam=\sixxm
    \textfont\ymfam=\elevenym
      \scriptfont\ymfam=\eightym
        \scriptscriptfont\ymfam=\sixym
   \fi
  \def\oldstyle{\fam\@ne\eleveni}%
  \def\boldstyle{\fam\mibfam\elevenmib}%
  \b@ls{13pt}\rm%
}

\def\fourteenpoint{
  \def\rm{\fam0\fourteenrm}%
  \textfont0\fourteenrm  \scriptfont0\tenrm  \scriptscriptfont0\sevenrm%
  \textfont1\fourteeni   \scriptfont1\teni   \scriptscriptfont1\seveni%
  \textfont2\fourteensy  \scriptfont2\tensy  \scriptscriptfont2\sevensy%
  \textfont\itfam=\fourteenit\def\it{\fam\itfam\fourteenit}%
  \ifprod@font
    \scriptfont\itfam=\tenit
      \scriptscriptfont\itfam=\sevenit
  \else
    \scriptfont\itfam=\fourteenit
      \scriptscriptfont\itfam=\fourteenit
  \fi
  \textfont\bffam=\fourteenbf%
    \scriptfont\bffam=\tenbf%
      \scriptscriptfont\bffam=\sevenbf%
  \def\bf{\fam\bffam\fourteenbf}%
  \textfont\slfam=\fourteensl\def\sl{\fam\slfam\fourteensl}%
  \ifprod@font
    \scriptfont\slfam=\tensl
      \scriptscriptfont\slfam=\sevensl
  \else
    \scriptfont\slfam=\fourteensl
      \scriptscriptfont\slfam=\fourteensl
  \fi
  \textfont\ttfam=\fourteentt\def\tt{\fam\ttfam\fourteentt}%
  \ifprod@font
    \scriptfont\ttfam=\tentt
      \scriptscriptfont\ttfam=\seventt
  \else
    \scriptfont\ttfam=\fourteentt
      \scriptscriptfont\ttfam=\fourteentt
  \fi
  \textfont\scfam=\fourteencsc\def\sc{\fam\scfam\fourteencsc}%
  \ifprod@font
    \scriptfont\scfam=\tencsc
      \scriptscriptfont\scfam=\sevencsc
  \else
    \scriptfont\scfam=\fourteencsc
      \scriptscriptfont\scfam=\fourteencsc
  \fi
  \textfont\sffam=\fourteensf\def\sf{\fam\sffam\fourteensf}%
  \ifprod@font
    \scriptfont\sffam=\tensf
      \scriptscriptfont\sffam=\sevensf
  \else
    \scriptfont\sffam=\fourteensf
      \scriptscriptfont\sffam=\fourteensf
  \fi
  \textfont\mibfam=\fourteenmib
    \scriptfont\mibfam=\tenmib
      \scriptscriptfont\mibfam=\sevenmib
  \textfont\sybfam=\fourteensyb
    \scriptfont\sybfam=\tensyb
      \scriptscriptfont\sybfam=\sevensyb
  \ifprod@font
    \textfont\xmfam=\fourteenxm
      \scriptfont\xmfam=\tenxm
        \scriptscriptfont\xmfam=\sevenxm
   \textfont\ymfam=\fourteenym
      \scriptfont\ymfam=\tenym
        \scriptscriptfont\ymfam=\sevenym
  \fi
  \def\oldstyle{\fam\@ne\fourteeni}%
  \def\boldstyle{\fam\mibfam\fourteenmib}%
  \b@ls{17pt}\rm%
}

\def\seventeenpoint{
  \def\rm{\fam0\seventeenrm}%
  \textfont0\seventeenrm  \scriptfont0\twelverm  \scriptscriptfont0\tenrm%
  \textfont1\seventeeni   \scriptfont1\twelvei   \scriptscriptfont1\teni%
  \textfont2\seventeensy  \scriptfont2\twelvesy  \scriptscriptfont2\tensy%
  \textfont\itfam=\seventeenit\def\it{\fam\itfam\seventeenit}%
  \ifprod@font
    \scriptfont\itfam=\twelveit
      \scriptscriptfont\itfam=\tenit
  \else
    \scriptfont\itfam=\seventeenit
      \scriptscriptfont\itfam=\seventeenit
  \fi
  \textfont\bffam=\seventeenbf%
    \scriptfont\bffam=\twelvebf%
      \scriptscriptfont\bffam=\tenbf%
  \def\bf{\fam\bffam\seventeenbf}%
  \textfont\slfam=\seventeensl\def\sl{\fam\slfam\seventeensl}%
  \ifprod@font
    \scriptfont\slfam=\twelvesl
      \scriptscriptfont\slfam=\tensl
  \else
    \scriptfont\slfam=\seventeensl
      \scriptscriptfont\slfam=\seventeensl
  \fi
  \textfont\ttfam=\seventeentt\def\tt{\fam\ttfam\seventeentt}%
  \ifprod@font
    \scriptfont\ttfam=\twelvett
      \scriptscriptfont\ttfam=\tentt
  \else
    \scriptfont\ttfam=\seventeentt
      \scriptscriptfont\ttfam=\seventeentt
  \fi
  \textfont\scfam=\seventeencsc\def\sc{\fam\scfam\seventeencsc}%
  \ifprod@font
    \scriptfont\scfam=\twelvecsc
      \scriptscriptfont\scfam=\tencsc
  \else
    \scriptfont\scfam=\seventeencsc
      \scriptscriptfont\scfam=\seventeencsc
  \fi
  \textfont\sffam=\seventeensf\def\sf{\fam\sffam\seventeensf}%
  \ifprod@font
    \scriptfont\sffam=\twelvesf
      \scriptscriptfont\sffam=\tensf
  \else
    \scriptfont\sffam=\seventeensf
      \scriptscriptfont\sffam=\seventeensf
  \fi
  \textfont\mibfam=\seventeenmib
    \scriptfont\mibfam=\twelvemib
      \scriptscriptfont\mibfam=\tenmib
  \textfont\sybfam=\seventeensyb
    \scriptfont\sybfam=\twelvesyb
      \scriptscriptfont\sybfam=\tensyb
  \ifprod@font
    \textfont\xmfam=\seventeenxm
      \scriptfont\xmfam=\twelvexm
        \scriptscriptfont\xmfam=\tenxm
    \textfont\ymfam=\seventeenym
      \scriptfont\ymfam=\twelveym
        \scriptscriptfont\ymfam=\tenym
  \fi
  \def\oldstyle{\fam\@ne\seventeeni}%
  \def\boldstyle{\fam\mibfam\seventeenmib}%
  \b@ls{20pt}\rm%
}

\lineskip=1pt      \normallineskip=\lineskip
\lineskiplimit=\z@ \normallineskiplimit=\lineskiplimit


\def\loadboldmathnames{%
  \mathchardef\balpha="0\mib0B
  \mathchardef\bbeta="0\mib0C
  \mathchardef\bgamma="0\mib0D
  \mathchardef\bdelta="0\mib0E
  \mathchardef\bepsilon="0\mib0F
  \mathchardef\bzeta="0\mib10
  \mathchardef\boldeta="0\mib11 
  \mathchardef\btheta="0\mib12
  \mathchardef\biota="0\mib13
  \mathchardef\bkappa="0\mib14
  \mathchardef\blambda="0\mib15
  \mathchardef\bmu="0\mib16
  \mathchardef\bnu="0\mib17
  \mathchardef\bxi="0\mib18
  \mathchardef\bpi="0\mib19
  \mathchardef\brho="0\mib1A
  \mathchardef\bsigma="0\mib1B
  \mathchardef\btau="0\mib1C
  \mathchardef\bupsilon="0\mib1D
  \mathchardef\bphi="0\mib1E
  \mathchardef\bchi="0\mib1F
  \mathchardef\bpsi="0\mib20
  \mathchardef\bomega="0\mib21
  \mathchardef\bvarepsilon="0\mib22
  \mathchardef\bvartheta="0\mib23
  \mathchardef\bvarpi="0\mib24
  \mathchardef\bvarrho="0\mib25
  \mathchardef\bvarsigma="0\mib26
  \mathchardef\bvarphi="0\mib27
  \mathchardef\baleph="0\syb40
  \mathchardef\bimath="0\mib7B
  \mathchardef\bjmath="0\mib7C
  \mathchardef\bell="0\mib60
  \mathchardef\bwp="0\mib7D
  \mathchardef\bRe="0\syb3C
  \mathchardef\bIm="0\syb3D
  \mathchardef\bpartial="0\mib40
  \mathchardef\binfty="0\syb31
  \mathchardef\bprime="0\syb30
  \mathchardef\bemptyset="0\syb3B
  \mathchardef\bnabla="0\syb72
  \mathchardef\btop="0\syb3E
  \mathchardef\bbot="0\syb3F
  \mathchardef\btriangle="0\syb34
  \mathchardef\bforall="0\syb38
  \mathchardef\bexists="0\syb39
  \mathchardef\bneg="0\syb3A
  \mathchardef\bflat="0\mib5B
  \mathchardef\bnatural="0\mib5C
  \mathchardef\bsharp="0\mib5D
  \mathchardef\bclubsuit="0\syb7C
  \mathchardef\bdiamondsuit="0\syb7D
  \mathchardef\bheartsuit="0\syb7E
  \mathchardef\bspadesuit="0\syb7F
  \mathchardef\bsmallint="1\syb73
  \mathchardef\btriangleleft="2\mib2F
  \mathchardef\btriangleright="2\mib2E
  \mathchardef\bbigtriangleup="2\syb34
  \mathchardef\bbigtriangledown="2\syb35
  \mathchardef\bwedge="2\syb5E
  \mathchardef\bvee="2\syb5F
  \mathchardef\bcap="2\syb5C
  \mathchardef\bcup="2\syb5B
  \mathchardef\bddagger="2\syb7A
  \mathchardef\bdagger="2\syb79
  \mathchardef\bsqcap="2\syb75
  \mathchardef\bsqcup="2\syb74
  \mathchardef\buplus="2\syb5D
  \mathchardef\bamalg="2\syb71
  \mathchardef\bdiamond="2\syb05
  \mathchardef\bbullet="2\syb0F
  \mathchardef\bwr="2\syb6F
  \mathchardef\bdiv="2\syb04
  \mathchardef\bodot="2\syb0C
  \mathchardef\boslash="2\syb0B
  \mathchardef\botimes="2\syb0A
  \mathchardef\bominus="2\syb09
  \mathchardef\boplus="2\syb08
  \mathchardef\bmp="2\syb07
  \mathchardef\bpm="2\syb06
  \mathchardef\bcirc="2\syb0E
  \mathchardef\bbigcirc="2\syb0D
  \mathchardef\bsetminus="2\syb6E
  \mathchardef\bcdot="2\syb01
  \mathchardef\bast="2\syb03
  \mathchardef\btimes="2\syb02
  \mathchardef\bstar="2\mib3F
  \mathchardef\bpropto="3\syb2F
  \mathchardef\bsqsubseteq="3\syb76
  \mathchardef\bsqsupseteq="3\syb77
  \mathchardef\bparallel="3\syb6B
  \mathchardef\bmid="3\syb6A
  \mathchardef\bdashv="3\syb61
  \mathchardef\bvdash="3\syb60
  \mathchardef\bnearrow="3\syb25
  \mathchardef\bsearrow="3\syb26
  \mathchardef\bnwarrow="3\syb2D
  \mathchardef\bswarrow="3\syb2E
  \mathchardef\bLeftrightarrow="3\syb2C
  \mathchardef\bLeftarrow="3\syb28
  \mathchardef\bRightarrow="3\syb29
  \mathchardef\bleq="3\syb14
  \mathchardef\bgeq="3\syb15
  \mathchardef\bsucc="3\syb1F
  \mathchardef\bprec="3\syb1E
  \mathchardef\bapprox="3\syb19
  \mathchardef\bsucceq="3\syb17
  \mathchardef\bpreceq="3\syb16
  \mathchardef\bsupset="3\syb1B
  \mathchardef\bsubset="3\syb1A
  \mathchardef\bsupseteq="3\syb13
  \mathchardef\bsubseteq="3\syb12
  \mathchardef\bin="3\syb32
  \mathchardef\bni="3\syb33
  \mathchardef\bgg="3\syb1D
  \mathchardef\bll="3\syb1C
  \mathchardef\bnot="3\syb36
  \mathchardef\bleftrightarrow="3\syb24
  \mathchardef\bleftarrow="3\syb20
  \mathchardef\brightarrow="3\syb21
  \mathchardef\bmapstochar="3\syb37
  \mathchardef\bsim="3\syb18
  \mathchardef\bsimeq="3\syb27
  \mathchardef\bperp="3\syb3F
  \mathchardef\bequiv="3\syb11
  \mathchardef\basymp="3\syb10
  \mathchardef\bsmile="3\mib5E
  \mathchardef\bfrown="3\mib5F
  \mathchardef\bleftharpoonup="3\mib28
  \mathchardef\bleftharpoondown="3\mib29
  \mathchardef\brightharpoonup="3\mib2A
  \mathchardef\brightharpoondown="3\mib2B
  \mathchardef\blhook="3\mib2C
  \mathchardef\brhook="3\mib2D
  \mathchardef\bldotp="6\mib3A
  \mathchardef\bcdotp="6\syb01
}


\def\la{\mathrel{\mathchoice {\vcenter{\offinterlineskip\halign{\hfil
$\displaystyle##$\hfil\cr<\cr\sim\cr}}}
{\vcenter{\offinterlineskip\halign{\hfil$\textstyle##$\hfil\cr
<\cr\sim\cr}}}
{\vcenter{\offinterlineskip\halign{\hfil$\scriptstyle##$\hfil\cr
<\cr\sim\cr}}}
{\vcenter{\offinterlineskip\halign{\hfil$\scriptscriptstyle##$\hfil\cr
<\cr\sim\cr}}}}}

\def\ga{\mathrel{\mathchoice {\vcenter{\offinterlineskip\halign{\hfil
$\displaystyle##$\hfil\cr>\cr\sim\cr}}}
{\vcenter{\offinterlineskip\halign{\hfil$\textstyle##$\hfil\cr
>\cr\sim\cr}}}
{\vcenter{\offinterlineskip\halign{\hfil$\scriptstyle##$\hfil\cr
>\cr\sim\cr}}}
{\vcenter{\offinterlineskip\halign{\hfil$\scriptscriptstyle##$\hfil\cr
>\cr\sim\cr}}}}}

\def\getsto{\mathrel{\mathchoice {\vcenter{\offinterlineskip
\halign{\hfil
$\displaystyle##$\hfil\cr\gets\cr\to\cr}}}
{\vcenter{\offinterlineskip\halign{\hfil$\textstyle##$\hfil\cr\gets
\cr\to\cr}}}
{\vcenter{\offinterlineskip\halign{\hfil$\scriptstyle##$\hfil\cr\gets
\cr\to\cr}}}
{\vcenter{\offinterlineskip\halign{\hfil$\scriptscriptstyle##$\hfil\cr
\gets\cr\to\cr}}}}}

\def\lid{\mathrel{\mathchoice {\vcenter{\offinterlineskip\halign{\hfil
$\displaystyle##$\hfil\cr<\cr\noalign{\vskip1.2pt}=\cr}}}
{\vcenter{\offinterlineskip\halign{\hfil$\textstyle##$\hfil\cr<\cr
\noalign{\vskip1.2pt}=\cr}}}
{\vcenter{\offinterlineskip\halign{\hfil$\scriptstyle##$\hfil\cr<\cr
\noalign{\vskip1pt}=\cr}}}
{\vcenter{\offinterlineskip\halign{\hfil$\scriptscriptstyle##$\hfil\cr
<\cr
\noalign{\vskip0.9pt}=\cr}}}}}

\def\gid{\mathrel{\mathchoice {\vcenter{\offinterlineskip\halign{\hfil
$\displaystyle##$\hfil\cr>\cr\noalign{\vskip1.2pt}=\cr}}}
{\vcenter{\offinterlineskip\halign{\hfil$\textstyle##$\hfil\cr>\cr
\noalign{\vskip1.2pt}=\cr}}}
{\vcenter{\offinterlineskip\halign{\hfil$\scriptstyle##$\hfil\cr>\cr
\noalign{\vskip1pt}=\cr}}}
{\vcenter{\offinterlineskip\halign{\hfil$\scriptscriptstyle##$\hfil\cr
>\cr
\noalign{\vskip0.9pt}=\cr}}}}}

\def\grole{\mathrel{\mathchoice {\vcenter{\offinterlineskip\halign{\hfil
$\displaystyle##$\hfil\cr>\cr\noalign{\vskip-1.5pt}<\cr}}}
{\vcenter{\offinterlineskip\halign{\hfil$\textstyle##$\hfil\cr
>\cr\noalign{\vskip-1.5pt}<\cr}}}
{\vcenter{\offinterlineskip\halign{\hfil$\scriptstyle##$\hfil\cr
>\cr\noalign{\vskip-1pt}<\cr}}}
{\vcenter{\offinterlineskip\halign{\hfil$\scriptscriptstyle##$\hfil\cr
>\cr\noalign{\vskip-0.5pt}<\cr}}}}}

\def\leogr{\mathrel{\mathchoice {\vcenter{\offinterlineskip\halign{\hfil
$\displaystyle##$\hfil\cr<\cr\noalign{\vskip-1.5pt}>\cr}}}
{\vcenter{\offinterlineskip\halign{\hfil$\textstyle##$\hfil\cr
<\cr\noalign{\vskip-1.5pt}>\cr}}}
{\vcenter{\offinterlineskip\halign{\hfil$\scriptstyle##$\hfil\cr
<\cr\noalign{\vskip-1pt}>\cr}}}
{\vcenter{\offinterlineskip\halign{\hfil$\scriptscriptstyle##$\hfil\cr
<\cr\noalign{\vskip-0.5pt}>\cr}}}}}

\def\loa{\mathrel{\mathchoice {\vcenter{\offinterlineskip\halign{\hfil
$\displaystyle##$\hfil\cr<\cr\approx\cr}}}
{\vcenter{\offinterlineskip\halign{\hfil$\textstyle##$\hfil\cr
<\cr\approx\cr}}}
{\vcenter{\offinterlineskip\halign{\hfil$\scriptstyle##$\hfil\cr
<\cr\approx\cr}}}
{\vcenter{\offinterlineskip\halign{\hfil$\scriptscriptstyle##$\hfil\cr
<\cr\approx\cr}}}}}

\def\goa{\mathrel{\mathchoice {\vcenter{\offinterlineskip\halign{\hfil
$\displaystyle##$\hfil\cr>\cr\approx\cr}}}
{\vcenter{\offinterlineskip\halign{\hfil$\textstyle##$\hfil\cr
>\cr\approx\cr}}}
{\vcenter{\offinterlineskip\halign{\hfil$\scriptstyle##$\hfil\cr
>\cr\approx\cr}}}
{\vcenter{\offinterlineskip\halign{\hfil$\scriptscriptstyle##$\hfil\cr
>\cr\approx\cr}}}}}

\def\diameter{{\ifmmode\mathchoice
{\ooalign{\hfil\hbox{$\displaystyle/$}\hfil\crcr
{\hbox{$\displaystyle\mathchar"20D$}}}}
{\ooalign{\hfil\hbox{$\textstyle/$}\hfil\crcr
{\hbox{$\textstyle\mathchar"20D$}}}}
{\ooalign{\hfil\hbox{$\scriptstyle/$}\hfil\crcr
{\hbox{$\scriptstyle\mathchar"20D$}}}}
{\ooalign{\hfil\hbox{$\scriptscriptstyle/$}\hfil\crcr
{\hbox{$\scriptscriptstyle\mathchar"20D$}}}}
\else{\ooalign{\hfil/\hfil\crcr\mathhexbox20D}}%
\fi}}

\def\sq{\ifmmode\squareforqed\else{\unskip\nobreak\hfil
\penalty50\hskip1em\null\nobreak\hfil\squareforqed
\parfillskip=0pt\finalhyphendemerits=0\endgraf}\fi}
\def\squareforqed{\hbox{\rlap{$\sqcap$}$\sqcup$}}


\def\bbbc{{\mathchoice {\setbox0=\hbox{$\displaystyle\rm C$}\hbox{\hbox
to0pt{\kern0.4\wd0\vrule height0.9\ht0\hss}\box0}}
{\setbox0=\hbox{$\textstyle\rm C$}\hbox{\hbox
to0pt{\kern0.4\wd0\vrule height0.9\ht0\hss}\box0}}
{\setbox0=\hbox{$\scriptstyle\rm C$}\hbox{\hbox
to0pt{\kern0.4\wd0\vrule height0.9\ht0\hss}\box0}}
{\setbox0=\hbox{$\scriptscriptstyle\rm C$}\hbox{\hbox
to0pt{\kern0.4\wd0\vrule height0.9\ht0\hss}\box0}}}}
\def\bbbq{{\mathchoice {\setbox0=\hbox{$\displaystyle\rm
Q$}\hbox{\raise
0.15\ht0\hbox to0pt{\kern0.4\wd0\vrule height0.8\ht0\hss}\box0}}
{\setbox0=\hbox{$\textstyle\rm Q$}\hbox{\raise
0.15\ht0\hbox to0pt{\kern0.4\wd0\vrule height0.8\ht0\hss}\box0}}
{\setbox0=\hbox{$\scriptstyle\rm Q$}\hbox{\raise
0.15\ht0\hbox to0pt{\kern0.4\wd0\vrule height0.7\ht0\hss}\box0}}
{\setbox0=\hbox{$\scriptscriptstyle\rm Q$}\hbox{\raise
0.15\ht0\hbox to0pt{\kern0.4\wd0\vrule height0.7\ht0\hss}\box0}}}}
\def\bbbt{{\mathchoice {\setbox0=\hbox{$\displaystyle\rm
T$}\hbox{\hbox to0pt{\kern0.3\wd0\vrule height0.9\ht0\hss}\box0}}
{\setbox0=\hbox{$\textstyle\rm T$}\hbox{\hbox
to0pt{\kern0.3\wd0\vrule height0.9\ht0\hss}\box0}}
{\setbox0=\hbox{$\scriptstyle\rm T$}\hbox{\hbox
to0pt{\kern0.3\wd0\vrule height0.9\ht0\hss}\box0}}
{\setbox0=\hbox{$\scriptscriptstyle\rm T$}\hbox{\hbox
to0pt{\kern0.3\wd0\vrule height0.9\ht0\hss}\box0}}}}
\def\bbbs{{\mathchoice
{\setbox0=\hbox{$\displaystyle     \rm S$}\hbox{\raise0.5\ht0\hbox
to0pt{\kern0.35\wd0\vrule height0.45\ht0\hss}\hbox
to0pt{\kern0.55\wd0\vrule height0.5\ht0\hss}\box0}}
{\setbox0=\hbox{$\textstyle        \rm S$}\hbox{\raise0.5\ht0\hbox
to0pt{\kern0.35\wd0\vrule height0.45\ht0\hss}\hbox
to0pt{\kern0.55\wd0\vrule height0.5\ht0\hss}\box0}}
{\setbox0=\hbox{$\scriptstyle      \rm S$}\hbox{\raise0.5\ht0\hbox
to0pt{\kern0.35\wd0\vrule height0.45\ht0\hss}\raise0.05\ht0\hbox
to0pt{\kern0.5\wd0\vrule height0.45\ht0\hss}\box0}}
{\setbox0=\hbox{$\scriptscriptstyle\rm S$}\hbox{\raise0.5\ht0\hbox
to0pt{\kern0.4\wd0\vrule height0.45\ht0\hss}\raise0.05\ht0\hbox
to0pt{\kern0.55\wd0\vrule height0.45\ht0\hss}\box0}}}}
\def\bbbz{{\mathchoice {\hbox{$\sf\textstyle Z\kern-0.4em Z$}}
{\hbox{$\sf\textstyle Z\kern-0.4em Z$}}
{\hbox{$\sf\scriptstyle Z\kern-0.3em Z$}}
{\hbox{$\sf\scriptscriptstyle Z\kern-0.2em Z$}}}}


\ifprod@font
  \mathchardef\la="3\@xm2E
  \mathchardef\getsto="3\@xm1C
  \mathchardef\lid="3\@xm35
  \mathchardef\grole="3\@xm3F
  \mathchardef\loa="3\@xm2F
  \mathchardef\ga="3\@xm26
  \mathchardef\gid="3\@xm3D
  \mathchardef\leogr="3\@xm37
  \mathchardef\goa="3\@xm27
  \mathchardef\sq="0\@xm03
%
%
\def\diameter{{%
  \ifmmode
    \mathchoice
    {\ooalign{\hfil\hbox{$\displaystyle/$}\hfil\crcr
    {\lower.2ex\hbox{$\displaystyle\mathchar"20D$}}}}%
    {\ooalign{\hfil\hbox{$\textstyle/$}\hfil\crcr
    {\lower.2ex\hbox{$\textstyle\mathchar"20D$}}}}%
    {\ooalign{\hfil\hbox{$\scriptstyle/$}\hfil\crcr
    {\lower.1ex\hbox{$\scriptstyle\mathchar"20D$}}}}%
    {\ooalign{\hfil\hbox{$\scriptscriptstyle/$}\hfil\crcr
    {\lower.1ex\hbox{$\scriptscriptstyle\mathchar"20D$}}}}%
  \else
    {\ooalign{\hfil/\hfil\crcr\lower.2ex\hbox{\mathhexbox20D}}}%
  \fi
}}
%
%

\def\bbbc{{\Bbb{C}}}
\def\bbbq{{\Bbb{Q}}}
\def\bbbt{{\Bbb{T}}}
\def\bbbs{{\Bbb{S}}}
\def\bbbz{{\Bbb{Z}}}
\fi


\ifprod@font
\mathchardef\boxdot="2\@xm00
\mathchardef\boxplus="2\@xm01
\mathchardef\boxtimes="2\@xm02
\mathchardef\square="0\@xm03
\mathchardef\blacksquare="0\@xm04
\mathchardef\centerdot="2\@xm05
\mathchardef\lozenge="0\@xm06
\mathchardef\blacklozenge="0\@xm07
\mathchardef\circlearrowright="3\@xm08
\mathchardef\circlearrowleft="3\@xm09
\mathchardef\rightleftharpoons="3\@xm0A
\mathchardef\leftrightharpoons="3\@xm0B
\mathchardef\boxminus="2\@xm0C
\mathchardef\Vdash="3\@xm0D
\mathchardef\Vvdash="3\@xm0E
\mathchardef\vDash="3\@xm0F
\mathchardef\twoheadrightarrow="3\@xm10
\mathchardef\twoheadleftarrow="3\@xm11
\mathchardef\leftleftarrows="3\@xm12
\mathchardef\rightrightarrows="3\@xm13
\mathchardef\upuparrows="3\@xm14
\mathchardef\downdownarrows="3\@xm15
\mathchardef\upharpoonright="3\@xm16

\mathchardef\downharpoonright="3\@xm17
\mathchardef\upharpoonleft="3\@xm18
\mathchardef\downharpoonleft="3\@xm19
\mathchardef\rightarrowtail="3\@xm1A
\mathchardef\leftarrowtail="3\@xm1B
\mathchardef\leftrightarrows="3\@xm1C
\mathchardef\rightleftarrows="3\@xm1D
\mathchardef\Lsh="3\@xm1E
\mathchardef\Rsh="3\@xm1F
\mathchardef\rightsquigarrow="3\@xm20
\mathchardef\leftrightsquigarrow="3\@xm21
\mathchardef\looparrowleft="3\@xm22
\mathchardef\looparrowright="3\@xm23
\mathchardef\circeq="3\@xm24
\mathchardef\succsim="3\@xm25
\mathchardef\gtrsim="3\@xm26
\mathchardef\gtrapprox="3\@xm27
\mathchardef\multimap="3\@xm28
\mathchardef\therefore="3\@xm29
\mathchardef\because="3\@xm2A
\mathchardef\doteqdot="3\@xm2B

\mathchardef\triangleq="3\@xm2C
\mathchardef\precsim="3\@xm2D
\mathchardef\lesssim="3\@xm2E
\mathchardef\lessapprox="3\@xm2F
\mathchardef\eqslantless="3\@xm30
\mathchardef\eqslantgtr="3\@xm31
\mathchardef\curlyeqprec="3\@xm32
\mathchardef\curlyeqsucc="3\@xm33
\mathchardef\preccurlyeq="3\@xm34
\mathchardef\leqq="3\@xm35
\mathchardef\leqslant="3\@xm36
\mathchardef\lessgtr="3\@xm37
\mathchardef\backprime="0\@xm38
\mathchardef\risingdotseq="3\@xm3A
\mathchardef\fallingdotseq="3\@xm3B
\mathchardef\succcurlyeq="3\@xm3C
\mathchardef\geqq="3\@xm3D
\mathchardef\geqslant="3\@xm3E
\mathchardef\gtrless="3\@xm3F
\mathchardef\sqsubset="3\@xm40
\mathchardef\sqsupset="3\@xm41
\mathchardef\vartriangleright="3\@xm42
\mathchardef\vartriangleleft="3\@xm43
\mathchardef\trianglerighteq="3\@xm44
\mathchardef\trianglelefteq="3\@xm45
\mathchardef\bigstar="0\@xm46
\mathchardef\between="3\@xm47
\mathchardef\blacktriangledown="0\@xm48
\mathchardef\blacktriangleright="3\@xm49
\mathchardef\blacktriangleleft="3\@xm4A
\mathchardef\vartriangle="0\@xm4D
\mathchardef\blacktriangle="0\@xm4E
\mathchardef\triangledown="0\@xm4F
\mathchardef\eqcirc="3\@xm50
\mathchardef\lesseqgtr="3\@xm51
\mathchardef\gtreqless="3\@xm52
\mathchardef\lesseqqgtr="3\@xm53
\mathchardef\gtreqqless="3\@xm54
\mathchardef\Rrightarrow="3\@xm56
\mathchardef\Lleftarrow="3\@xm57
\mathchardef\veebar="2\@xm59
\mathchardef\barwedge="2\@xm5A
\mathchardef\doublebarwedge="2\@xm5B
\mathchardef\angle="0\@xm5C
\mathchardef\measuredangle="0\@xm5D
\mathchardef\sphericalangle="0\@xm5E
\mathchardef\varpropto="3\@xm5F
\mathchardef\smallsmile="3\@xm60
\mathchardef\smallfrown="3\@xm61
\mathchardef\Subset="3\@xm62
\mathchardef\Supset="3\@xm63
\mathchardef\Cup="2\@xm64

\mathchardef\Cap="2\@xm65

\mathchardef\curlywedge="2\@xm66
\mathchardef\curlyvee="2\@xm67
\mathchardef\leftthreetimes="2\@xm68
\mathchardef\rightthreetimes="2\@xm69
\mathchardef\subseteqq="3\@xm6A
\mathchardef\supseteqq="3\@xm6B
\mathchardef\bumpeq="3\@xm6C
\mathchardef\Bumpeq="3\@xm6D
\mathchardef\lll="3\@xm6E

\mathchardef\ggg="3\@xm6F

\mathchardef\circledS="0\@xm73
\mathchardef\pitchfork="3\@xm74
\mathchardef\dotplus="2\@xm75
\mathchardef\backsim="3\@xm76
\mathchardef\backsimeq="3\@xm77
\mathchardef\complement="0\@xm7B
\mathchardef\intercal="2\@xm7C
\mathchardef\circledcirc="2\@xm7D
\mathchardef\circledast="2\@xm7E
\mathchardef\circleddash="2\@xm7F
\def\ulcorner{\delimiter"4\@xm70\@xm70 }
\def\urcorner{\delimiter"5\@xm71\@xm71 }
\def\llcorner{\delimiter"4\@xm78\@xm78 }
\def\lrcorner{\delimiter"5\@xm79\@xm79 }
\def\yen{\mathhexbox\@xm55 }
\def\checkmark{\mathhexbox\@xm58 }
\def\circledR{\mathhexbox\@xm72 }
\def\maltese{\mathhexbox\@xm7A }
\mathchardef\lvertneqq="3\@ym00
\mathchardef\gvertneqq="3\@ym01
\mathchardef\nleq="3\@ym02
\mathchardef\ngeq="3\@ym03
\mathchardef\nless="3\@ym04
\mathchardef\ngtr="3\@ym05
\mathchardef\nprec="3\@ym06
\mathchardef\nsucc="3\@ym07
\mathchardef\lneqq="3\@ym08
\mathchardef\gneqq="3\@ym09
\mathchardef\nleqslant="3\@ym0A
\mathchardef\ngeqslant="3\@ym0B
\mathchardef\lneq="3\@ym0C
\mathchardef\gneq="3\@ym0D
\mathchardef\npreceq="3\@ym0E
\mathchardef\nsucceq="3\@ym0F
\mathchardef\precnsim="3\@ym10
\mathchardef\succnsim="3\@ym11
\mathchardef\lnsim="3\@ym12
\mathchardef\gnsim="3\@ym13
\mathchardef\nleqq="3\@ym14
\mathchardef\ngeqq="3\@ym15
\mathchardef\precneqq="3\@ym16
\mathchardef\succneqq="3\@ym17
\mathchardef\precnapprox="3\@ym18
\mathchardef\succnapprox="3\@ym19
\mathchardef\lnapprox="3\@ym1A
\mathchardef\gnapprox="3\@ym1B
\mathchardef\nsim="3\@ym1C
\mathchardef\ncong="3\@ym1D

\mathchardef\varsubsetneq="3\@ym20
\mathchardef\varsupsetneq="3\@ym21
\mathchardef\nsubseteqq="3\@ym22
\mathchardef\nsupseteqq="3\@ym23
\mathchardef\subsetneqq="3\@ym24
\mathchardef\supsetneqq="3\@ym25
\mathchardef\varsubsetneqq="3\@ym26
\mathchardef\varsupsetneqq="3\@ym27
\mathchardef\subsetneq="3\@ym28
\mathchardef\supsetneq="3\@ym29
\mathchardef\nsubseteq="3\@ym2A
\mathchardef\nsupseteq="3\@ym2B
\mathchardef\nparallel="3\@ym2C
\mathchardef\nmid="3\@ym2D
\mathchardef\nshortmid="3\@ym2E
\mathchardef\nshortparallel="3\@ym2F
\mathchardef\nvdash="3\@ym30
\mathchardef\nVdash="3\@ym31
\mathchardef\nvDash="3\@ym32
\mathchardef\nVDash="3\@ym33
\mathchardef\ntrianglerighteq="3\@ym34
\mathchardef\ntrianglelefteq="3\@ym35
\mathchardef\ntriangleleft="3\@ym36
\mathchardef\ntriangleright="3\@ym37
\mathchardef\nleftarrow="3\@ym38
\mathchardef\nrightarrow="3\@ym39
\mathchardef\nLeftarrow="3\@ym3A
\mathchardef\nRightarrow="3\@ym3B
\mathchardef\nLeftrightarrow="3\@ym3C
\mathchardef\nleftrightarrow="3\@ym3D
\mathchardef\divideontimes="2\@ym3E
\mathchardef\varnothing="0\@ym3F
\mathchardef\nexists="0\@ym40
\mathchardef\mho="0\@ym66
\mathchardef\eth="0\@ym67
\mathchardef\eqsim="3\@ym68
\mathchardef\beth="0\@ym69
\mathchardef\gimel="0\@ym6A
\mathchardef\daleth="0\@ym6B
\mathchardef\lessdot="3\@ym6C
\mathchardef\gtrdot="3\@ym6D
\mathchardef\ltimes="2\@ym6E
\mathchardef\rtimes="2\@ym6F
\mathchardef\shortmid="3\@ym70
\mathchardef\shortparallel="3\@ym71
\mathchardef\smallsetminus="2\@ym72
\mathchardef\thicksim="3\@ym73
\mathchardef\thickapprox="3\@ym74
\mathchardef\approxeq="3\@ym75
\mathchardef\succapprox="3\@ym76
\mathchardef\precapprox="3\@ym77
\mathchardef\curvearrowleft="3\@ym78
\mathchardef\curvearrowright="3\@ym79
\mathchardef\digamma="0\@ym7A
\mathchardef\varkappa="0\@ym7B
\mathchardef\hslash="0\@ym7D
\mathchardef\hbar="0\@ym7E
\mathchardef\backepsilon="3\@ym7F


\def\Bbb{\ifmmode\let\next\Bbb@\else
\def\next{\errmessage{Use \string\Bbb\space only in math mode}}\fi\next}
\def\Bbb@#1{{\Bbb@@{#1}}}
\def\Bbb@@#1{\fam\ymfam#1}
\fi


\def\Nulle{0} 
\def\Afe{1}   
\def\Hae{2}   
\def\Hbe{3}   
\def\Hce{4}   
\def\Hde{5}   


\newcount\LastMac       \LastMac=\Nulle

\newskip\half      \half=5.5pt plus 1.5pt minus 2.25pt
\newskip\one       \one=11pt plus 3pt minus 5.5pt
\newskip\onehalf   \onehalf=16.5pt plus 5.5pt minus 8.25pt
\newskip\two       \two=22pt plus 5.5pt minus 11pt

\def\Half{\addvspace{\half}}
\def\One{\addvspace{\one}}
\def\OneHalf{\addvspace{\onehalf}}
\def\Two{\addvspace{\two}}


\def\Raggedright{
  \rightskip=\z@ plus \hsize\relax
}

\def\Fullout{
  \rightskip=\z@\relax
}

\def\Hang#1#2{
  \hangindent=#1%
  \hangafter=#2\relax
}


\newif\ifsp@page
\def\pagestyle#1{\csname ps@#1\endcsname}
\def\thispagestyle#1{\global\sp@pagetrue\gdef\sp@type{#1}}

\def\ps@titlepage{%
  \def\@oddhead{\eightpoint\noindent \the\CatchLine
    \ifprod@font\else\qquad Printed\ \today\fi \hfil}%
  \let\@evenhead=\@oddhead
}

\def\ps@headings{%
  \def\@oddhead{\elevenpoint\it\noindent
    \hfill\the\RightHeader\hskip1.5em\rm\folio}%
  \def\@evenhead{\elevenpoint\noindent
    \folio\hskip1.5em\it\the\LeftHeader\hfill}%
}

\def\ps@plate{%
  \def\@oddhead{\eightpoint\noindent\plt@cap\hfil}%
  \def\@evenhead{\eightpoint\noindent\plt@cap\hfil}%
}



\def\title#1{
  \bgroup
    \vbox to 8pt{\vss}%
    \seventeenpoint
    \Raggedright
    \noindent \strut{\bf #1}\par
  \egroup
}

\def\author#1{
  \bgroup
    \ifnum\LastMac=\Afe \OneHalf\else \vskip 21pt\fi
    \fourteenpoint
    \Raggedright
    \noindent \strut #1\par
    \vskip 3pt%
  \egroup
}

\def\affiliation#1{
  \bgroup
    \vskip -4pt%
    \eightpoint
    \Raggedright
    \noindent \strut {\it #1}\par
  \egroup
  \LastMac=\Afe\relax
}

\def\acceptedline#1{
  \bgroup
    \Two
    \eightpoint
    \Raggedright
    \noindent \strut #1\par
  \egroup
}

\long\def\abstract#1{%
  \bgroup
    \vskip 20pt%
    \everypar{\Hang{11pc}{0}}%
    \noindent{\ninebf ABSTRACT}\par
    \tenpoint
    \Fullout
    \noindent #1\par
  \egroup
}

\long\def\keywords#1{
  \bgroup
    \Half
    \everypar{\Hang{11pc}{0}}%
    \tenpoint
    \Fullout
    \noindent\hbox{\bf Key words:}\ #1\par
  \egroup
}


\def\maketitle{%
  \EndOpening
  \ifsinglecol \else \MakePage\fi
}



\def\Autonumber{
  \global\AutoNumbertrue  
}

\newif\ifAutoNumber \AutoNumberfalse
\newcount\Sec        
\newcount\SecSec
\newcount\SecSecSec

\Sec=\z@

\def\:{\let\@sptoken= } \:  
\def\:{\@xifnch} \expandafter\def\: {\futurelet\@tempc\@ifnch}

\def\@ifnextchar#1#2#3{%
  \let\@tempMACe #1%
  \def\@tempMACa{#2}%
  \def\@tempMACb{#3}%
  \futurelet \@tempMACc\@ifnch%
}

\def\@ifnch{%
\ifx \@tempMACc \@sptoken%
  \let\@tempMACd\@xifnch%
\else%
  \ifx \@tempMACc \@tempMACe%
    \let\@tempMACd\@tempMACa%
  \else%
    \let\@tempMACd\@tempMACb%
  \fi%
\fi%
\@tempMACd%
}

\def\@ifstar#1#2{\@ifnextchar *{\def\@tempMACa*{#1}\@tempMACa}{#2}}

\newskip\@tempskipb

\def\addvspace#1{%
  \ifvmode\else \endgraf\fi%
  \ifdim\lastskip=\z@%
    \vskip #1\relax%
  \else%
    \@tempskipb#1\relax\@xaddvskip%
  \fi%
}

\def\@xaddvskip{%
  \ifdim\lastskip<\@tempskipb%
    \vskip-\lastskip%
    \vskip\@tempskipb\relax%
  \else%
    \ifdim\@tempskipb<\z@%
      \ifdim\lastskip<\z@ \else%
        \advance\@tempskipb\lastskip%
        \vskip-\lastskip\vskip\@tempskipb%
      \fi%
    \fi%
  \fi%
}

\newskip\@tmpSKIP

\def\addpen#1{%
  \ifvmode
    \if@nobreak
    \else
      \ifdim\lastskip=\z@
        \penalty#1\relax
      \else
        \@tmpSKIP=\lastskip
        \vskip -\lastskip
        \penalty#1\vskip\@tmpSKIP
      \fi
    \fi
  \fi
}

\newcount\@clubpen   \@clubpen=\clubpenalty
\newif\if@nobreak    \@nobreakfalse

\def\@noafterindent{%
  \global\@nobreaktrue
  \everypar{\if@nobreak
              \global\@nobreakfalse
              \clubpenalty \@M
              {\setbox\z@\lastbox}%
              \LastMac=\Nulle\relax%
            \else
              \clubpenalty \@clubpen
              \everypar{}%
            \fi}
}

\newcount\gds@cbrk   \gds@cbrk=-300

\def\@nohdbrk{\interlinepenalty \@M\relax}

\let\@par=\par
\def\@restorepar{\def\par{\@par}}

\newif\if@endpe   \@endpefalse
 
\def\@doendpe{\@endpetrue \@nobreakfalse \LastMac=\Nulle\relax%
     \def\par{\@restorepar\everypar{}\par\@endpefalse}%
              \everypar{\setbox\z@\lastbox\everypar{}\@endpefalse}%
}

\def\section{\@ifstar{\@ssection}{\@section}}

\def\@section#1{
  \if@nobreak
    \everypar{}%
    \ifnum\LastMac=\Hae \addvspace{\half}\fi
  \else
    \addpen{\gds@cbrk}%
    \addvspace{\two}%
  \fi
  \bgroup
    \ninepoint\bf
    \Raggedright
    \ifAutoNumber
      \global\advance\Sec \@ne
      \noindent\@nohdbrk\number\Sec\hskip 1pc \uppercase{#1}\par
      \global\SecSec=\z@
    \else
      \noindent\@nohdbrk\uppercase{#1}\par
    \fi
  \egroup
  \nobreak
  \vskip\half
  \nobreak
  \@noafterindent
  \LastMac=\Hae\relax
}

\def\@ssection#1{
  \if@nobreak
    \everypar{}%
    \ifnum\LastMac=\Hae \addvspace{\half}\fi
  \else
    \addpen{\gds@cbrk}%
    \addvspace{\two}%
  \fi
  \bgroup
    \ninepoint\bf
    \Raggedright
    \noindent\@nohdbrk\uppercase{#1}\par
  \egroup
  \nobreak
  \vskip\half
  \nobreak
  \@noafterindent
  \LastMac=\Hae\relax
}

\def\subsection#1{
  \if@nobreak
    \everypar{}%
    \ifnum\LastMac=\Hae \addvspace{1pt plus 1pt minus .5pt}\fi
  \else
    \addpen{\gds@cbrk}%
    \addvspace{\onehalf}%
  \fi
  \bgroup
    \ninepoint\bf
    \Raggedright
    \ifAutoNumber
      \global\advance\SecSec \@ne
      \noindent\@nohdbrk\number\Sec.\number\SecSec \hskip 1pc\relax #1\par
      \global\SecSecSec=\z@
    \else
      \noindent\@nohdbrk #1\par
    \fi
  \egroup
  \nobreak
  \vskip\half
  \nobreak
  \@noafterindent
  \LastMac=\Hbe\relax
}

\def\subsubsection#1{
  \if@nobreak
    \everypar{}%
    \ifnum\LastMac=\Hbe \addvspace{1pt plus 1pt minus .5pt}\fi
  \else
    \addpen{\gds@cbrk}%
    \addvspace{\onehalf}%
  \fi
  \bgroup
    \ninepoint\it
    \Raggedright
    \ifAutoNumber
      \global\advance\SecSecSec \@ne
      \noindent\@nohdbrk\number\Sec.\number\SecSec.\number\SecSecSec
        \hskip 1pc\relax #1\par
    \else
      \noindent\@nohdbrk #1\par
    \fi
  \egroup
  \nobreak
  \vskip\half
  \nobreak
  \@noafterindent
  \LastMac=\Hce\relax
}

\def\paragraph#1{
  \if@nobreak
    \everypar{}%
  \else
    \addpen{\gds@cbrk}%
    \addvspace{\one}%
  \fi%
  \bgroup%
    \ninepoint\it
    \noindent #1\ \nobreak%
  \egroup
  \LastMac=\Hde\relax
  \ignorespaces
}




\def\beginlist{%
  \par\if@nobreak \else\addvspace{\half}\fi%
  \bgroup%
    \ninepoint
    \let\item=\list@item%
}

\def\list@item{%
  \par\noindent\hskip 1em\relax%
  \ignorespaces%
}

\def\endlist{\par\egroup\addvspace{\half}\@doendpe}


\def\beginrefs{%
  \par
  \bgroup
    \eightpoint
    \Raggedright
    \let\bibitem=\bib@item
}

\def\bib@item{%
  \par\parindent=1.5em\Hang{1.5em}{1}%
  \everypar={\Hang{1.5em}{1}\ignorespaces}%
  \noindent\ignorespaces
}

\def\endrefs{\par\egroup\@doendpe}


\newtoks\CatchLine

\def\@journal{Mon.\ Not.\ R.\ Astron.\ Soc.\ }  
\def\@pubyear{1994}        
\def\@pagerange{000--000}  
\def\@volume{000}          
\def\@microfiche{}         %

\def\pubyear#1{\gdef\@pubyear{#1}\@makecatchline}
\def\pagerange#1{\gdef\@pagerange{#1}\@makecatchline}
\def\volume#1{\gdef\@volume{#1}\@makecatchline}
\def\microfiche#1{\gdef\@microfiche{and Microfiche\ #1}\@makecatchline}

\def\@makecatchline{%
  \global\CatchLine{%
    {\rm \@journal {\bf \@volume},\ \@pagerange\ (\@pubyear)\ \@microfiche}}%
}

\@makecatchline 

\newtoks\LeftHeader
\def\shortauthor#1{
  \global\LeftHeader{#1}%
}

\newtoks\RightHeader

\def\PageHead{
  \begingroup
    \ifsp@page
      \csname ps@\sp@type\endcsname
      \global\sp@pagefalse
    \fi
    \ifodd\pageno
      \let\the@head=\@oddhead
    \else
      \let\the@head=\@evenhead
    \fi
    \vbox to \z@{\vskip-22.5\p@%
      \hbox to \PageWidth{\vbox to8.5\p@{}%
        \the@head
      }%
    \vss}%
  \endgroup
  \nointerlineskip
}

\def\today{%
  \number\day\space
  \ifcase\month\or January\or February\or March\or April\or May\or June\or
    July\or August\or September\or October\or November\or December\fi
  \space\number\year%
}

\def\PageFoot{} 

\def\authorcomment#1{%
  \gdef\PageFoot{%
    \nointerlineskip%
    \vbox to 22pt{\vfil%
      \hbox to \PageWidth{\elevenpoint\noindent \hfil #1 \hfil}}%
  }%
}


\newif\ifplate@page
\newbox\plt@box

\def\beginplatepage{%
  \let\plate=\plate@head
  \let\caption=\fig@caption
  \global\setbox\plt@box=\vbox\bgroup
  \TEMPDIMEN=\PageWidth 
  \hsize=\PageWidth\relax
}

\def\endplatepage{\par\egroup\global\plate@pagetrue}
\def\plate@head#1{\gdef\plt@cap{#1}}


\def\letters{%
  \gdef\folio{\ifnum\pageno<\z@ L\romannumeral-\pageno
    \else L\number\pageno \fi}%
}


\everydisplay{\displaysetup}

\newif\ifeqno
\newif\ifleqno

\def\displaysetup#1$${%
 \displaytest#1\eqno\eqno\displaytest
}

\def\displaytest#1\eqno#2\eqno#3\displaytest{%
 \if!#3!\ldisplaytest#1\leqno\leqno\ldisplaytest
 \else\eqnotrue\leqnofalse\def\eqn{#2}\def\eq{#1}\fi
 \generaldisplay$$}

\def\ldisplaytest#1\leqno#2\leqno#3\ldisplaytest{%
 \def\eq{#1}%
 \if!#3!\eqnofalse\else\eqnotrue\leqnotrue
  \def\eqn{#2}\fi}

\def\generaldisplay{%
\ifeqno \ifleqno 
   \hbox to \hsize{\noindent
     $\displaystyle\eq$\hfil$\displaystyle\eqn$}
  \else
    \hbox to \hsize{\noindent
     $\displaystyle\eq$\hfil$\displaystyle\eqn$}
  \fi
 \else
 \hbox to \hsize{\vbox{\noindent
  $\displaystyle\eq$\hfil}}
 \fi
}


\def\@notice{%
  \par\Two%
  \noindent{\b@ls{11pt}\ninerm This paper has been produced using the
    Blackwell Scientific Publications \TeX\ macros.\par}%
}

\outer\def\bye{\@notice\par\vfill\supereject\end}


\def\start@mess{%
  Monthly notices of the RAS journal style (\@typeface)\space
    v\@version,\space \@verdate.%
}

\everyjob{\Warn{\start@mess}}



\newif\if@debug \@debugfalse  

\def\Print#1{\if@debug\immediate\write16{#1}\else \fi}
\def\Warn#1{\immediate\write16{#1}}
\def\wlog#1{}

\newcount\Iteration 

\def\Single{0} \def\Double{1}                 
\def\Figure{0} \def\Table{1}                  

\def\InStack{0}  
\def\InZoneA{1}
\def\InZoneB{2}
\def\InZoneC{3}

\newcount\TEMPCOUNT 
\newdimen\TEMPDIMEN 
\newbox\TEMPBOX     
\newbox\VOIDBOX     

\newcount\LengthOfStack 
\newcount\MaxItems      
\newcount\StackPointer
\newcount\Point         
\newcount\NextFigure    
\newcount\NextTable     
\newcount\NextItem      

\newcount\StatusStack   
\newcount\NumStack      
\newcount\TypeStack     
\newcount\SpanStack     
\newcount\BoxStack      

\newcount\ItemSTATUS    
\newcount\ItemNUMBER    
\newcount\ItemTYPE      
\newcount\ItemSPAN      
\newbox\ItemBOX         
\newdimen\ItemSIZE      

\newdimen\PageHeight    
\newdimen\TextLeading   
\newdimen\Feathering    
\newcount\LinesPerPage  
\newdimen\ColumnWidth   
\newdimen\ColumnGap     
\newdimen\PageWidth     
\newdimen\BodgeHeight   
\newcount\Leading       

\newdimen\ZoneBSize  
\newdimen\TextSize   
\newbox\ZoneABOX     
\newbox\ZoneBBOX     
\newbox\ZoneCBOX     

\newif\ifFirstSingleItem
\newif\ifFirstZoneA
\newif\ifMakePageInComplete
\newif\ifMoreFigures \MoreFiguresfalse 
\newif\ifMoreTables  \MoreTablesfalse  

\newif\ifFigInZoneB 
\newif\ifFigInZoneC 
\newif\ifTabInZoneB 
\newif\ifTabInZoneC

\newif\ifZoneAFullPage

\newbox\MidBOX    
\newbox\LeftBOX
\newbox\RightBOX
\newbox\PageBOX   

\newif\ifLeftCOL  
\LeftCOLtrue

\newdimen\ZoneBAdjust

\newcount\ItemFits
\def\Yes{1}
\def\No{2}


\MaxItems=15
\NextFigure=\z@        
\NextTable=\@ne

\BodgeHeight=6pt
\TextLeading=11pt    
\Leading=11
\Feathering=\z@      
\LinesPerPage=61     
\topskip=\TextLeading
\ColumnWidth=20pc    
\ColumnGap=2pc       

\newskip\ItemSepamount  
\ItemSepamount=\TextLeading plus \TextLeading minus 4pt

\parskip=\z@ plus .1pt
\parindent=18pt
\widowpenalty=\z@
\clubpenalty=10000
\tolerance=1500
\hbadness=1500
\abovedisplayskip=6pt plus 2pt minus 2pt
\belowdisplayskip=6pt plus 2pt minus 2pt
\abovedisplayshortskip=6pt plus 2pt minus 2pt
\belowdisplayshortskip=6pt plus 2pt minus 2pt

\ninepoint 


\PageHeight=682pt

\PageWidth=2\ColumnWidth
\advance\PageWidth by \ColumnGap

\pagestyle{headings}




\newcount\DUMMY \StatusStack=\allocationnumber
\newcount\DUMMY \newcount\DUMMY \newcount\DUMMY 
\newcount\DUMMY \newcount\DUMMY \newcount\DUMMY 
\newcount\DUMMY \newcount\DUMMY \newcount\DUMMY
\newcount\DUMMY \newcount\DUMMY \newcount\DUMMY 
\newcount\DUMMY \newcount\DUMMY \newcount\DUMMY

\newcount\DUMMY \NumStack=\allocationnumber
\newcount\DUMMY \newcount\DUMMY \newcount\DUMMY 
\newcount\DUMMY \newcount\DUMMY \newcount\DUMMY 
\newcount\DUMMY \newcount\DUMMY \newcount\DUMMY 
\newcount\DUMMY \newcount\DUMMY \newcount\DUMMY 
\newcount\DUMMY \newcount\DUMMY \newcount\DUMMY

\newcount\DUMMY \TypeStack=\allocationnumber
\newcount\DUMMY \newcount\DUMMY \newcount\DUMMY 
\newcount\DUMMY \newcount\DUMMY \newcount\DUMMY 
\newcount\DUMMY \newcount\DUMMY \newcount\DUMMY 
\newcount\DUMMY \newcount\DUMMY \newcount\DUMMY 
\newcount\DUMMY \newcount\DUMMY \newcount\DUMMY

\newcount\DUMMY \SpanStack=\allocationnumber
\newcount\DUMMY \newcount\DUMMY \newcount\DUMMY 
\newcount\DUMMY \newcount\DUMMY \newcount\DUMMY 
\newcount\DUMMY \newcount\DUMMY \newcount\DUMMY 
\newcount\DUMMY \newcount\DUMMY \newcount\DUMMY 
\newcount\DUMMY \newcount\DUMMY \newcount\DUMMY

\newbox\DUMMY   \BoxStack=\allocationnumber
\newbox\DUMMY   \newbox\DUMMY \newbox\DUMMY 
\newbox\DUMMY   \newbox\DUMMY \newbox\DUMMY 
\newbox\DUMMY   \newbox\DUMMY \newbox\DUMMY 
\newbox\DUMMY   \newbox\DUMMY \newbox\DUMMY 
\newbox\DUMMY   \newbox\DUMMY \newbox\DUMMY

\def\wlog{\immediate\write\m@ne}


\def\GetItemAll#1{%
 \GetItemSTATUS{#1}
 \GetItemNUMBER{#1}
 \GetItemTYPE{#1}
 \GetItemSPAN{#1}
 \GetItemBOX{#1}
}

\def\GetItemSTATUS#1{%
 \Point=\StatusStack
 \advance\Point by #1
 \global\ItemSTATUS=\count\Point
}

\def\GetItemNUMBER#1{%
 \Point=\NumStack
 \advance\Point by #1
 \global\ItemNUMBER=\count\Point
}

\def\GetItemTYPE#1{%
 \Point=\TypeStack
 \advance\Point by #1
 \global\ItemTYPE=\count\Point
}

\def\GetItemSPAN#1{%
 \Point\SpanStack
 \advance\Point by #1
 \global\ItemSPAN=\count\Point
}

\def\GetItemBOX#1{%
 \Point=\BoxStack
 \advance\Point by #1
 \global\setbox\ItemBOX=\vbox{\copy\Point}
 \global\ItemSIZE=\ht\ItemBOX
 \global\advance\ItemSIZE by \dp\ItemBOX
 \TEMPCOUNT=\ItemSIZE
 \divide\TEMPCOUNT by \Leading
 \divide\TEMPCOUNT by 65536
 \advance\TEMPCOUNT \@ne
 \ItemSIZE=\TEMPCOUNT pt
 \global\multiply\ItemSIZE by \Leading
}


\def\JoinStack{%
 \ifnum\LengthOfStack=\MaxItems 
  \Warn{WARNING: Stack is full...some items will be lost!}
 \else
  \Point=\StatusStack
  \advance\Point by \LengthOfStack
  \global\count\Point=\ItemSTATUS
  \Point=\NumStack
  \advance\Point by \LengthOfStack
  \global\count\Point=\ItemNUMBER
  \Point=\TypeStack
  \advance\Point by \LengthOfStack
  \global\count\Point=\ItemTYPE
  \Point\SpanStack
  \advance\Point by \LengthOfStack
  \global\count\Point=\ItemSPAN
  \Point=\BoxStack
  \advance\Point by \LengthOfStack
  \global\setbox\Point=\vbox{\copy\ItemBOX}
  \global\advance\LengthOfStack \@ne
  \ifnum\ItemTYPE=\Figure 
   \global\MoreFigurestrue
  \else
   \global\MoreTablestrue
  \fi
 \fi
}


\def\LeaveStack#1{%
 {\Iteration=#1
 \loop
 \ifnum\Iteration<\LengthOfStack
  \advance\Iteration \@ne
  \GetItemSTATUS{\Iteration}
   \advance\Point by \m@ne
   \global\count\Point=\ItemSTATUS
  \GetItemNUMBER{\Iteration}
   \advance\Point by \m@ne
   \global\count\Point=\ItemNUMBER
  \GetItemTYPE{\Iteration}
   \advance\Point by \m@ne
   \global\count\Point=\ItemTYPE
  \GetItemSPAN{\Iteration}
   \advance\Point by \m@ne
   \global\count\Point=\ItemSPAN
  \GetItemBOX{\Iteration}
   \advance\Point by \m@ne
   \global\setbox\Point=\vbox{\copy\ItemBOX}
 \repeat}
 \global\advance\LengthOfStack by \m@ne
}


\newif\ifStackNotClean

\def\CleanStack{%
 \StackNotCleantrue
 {\Iteration=\z@
  \loop
   \ifStackNotClean
    \GetItemSTATUS{\Iteration}
    \ifnum\ItemSTATUS=\InStack
     \advance\Iteration \@ne
     \else
      \LeaveStack{\Iteration}
    \fi
   \ifnum\LengthOfStack<\Iteration
    \StackNotCleanfalse
   \fi
 \repeat}
}


\def\FindItem#1#2{%
 \global\StackPointer=\m@ne 
 {\Iteration=\z@
  \loop
  \ifnum\Iteration<\LengthOfStack
   \GetItemSTATUS{\Iteration}
   \ifnum\ItemSTATUS=\InStack
    \GetItemTYPE{\Iteration}
    \ifnum\ItemTYPE=#1
     \GetItemNUMBER{\Iteration}
     \ifnum\ItemNUMBER=#2
      \global\StackPointer=\Iteration
      \Iteration=\LengthOfStack 
     \fi
    \fi
   \fi
  \advance\Iteration \@ne
 \repeat}
}


\def\FindNext{%
 \global\StackPointer=\m@ne 
 {\Iteration=\z@
  \loop
  \ifnum\Iteration<\LengthOfStack
   \GetItemSTATUS{\Iteration}
   \ifnum\ItemSTATUS=\InStack
    \GetItemTYPE{\Iteration}
   \ifnum\ItemTYPE=\Figure
    \ifMoreFigures
      \global\NextItem=\Figure
      \global\StackPointer=\Iteration
      \Iteration=\LengthOfStack 
    \fi
   \fi
   \ifnum\ItemTYPE=\Table
    \ifMoreTables
      \global\NextItem=\Table
      \global\StackPointer=\Iteration
      \Iteration=\LengthOfStack 
    \fi
   \fi
  \fi
  \advance\Iteration \@ne
 \repeat}
}


\def\ChangeStatus#1#2{%
 \Point=\StatusStack
 \advance\Point by #1
 \global\count\Point=#2
}



\def\Zone{\InZoneA}

\ZoneBAdjust=\z@

\def\MakePage{
 \global\ZoneBSize=\PageHeight
 \global\TextSize=\ZoneBSize
 \global\ZoneAFullPagefalse
 \global\topskip=\TextLeading
 \MakePageInCompletetrue
 \MoreFigurestrue
 \MoreTablestrue
 \FigInZoneBfalse
 \FigInZoneCfalse
 \TabInZoneBfalse
 \TabInZoneCfalse
 \global\FirstSingleItemtrue
 \global\FirstZoneAtrue
 \global\setbox\ZoneABOX=\box\VOIDBOX
 \global\setbox\ZoneBBOX=\box\VOIDBOX
 \global\setbox\ZoneCBOX=\box\VOIDBOX
 \loop
  \ifMakePageInComplete
 \FindNext
 \ifnum\StackPointer=\m@ne
  \NextItem=\m@ne
  \MoreFiguresfalse
  \MoreTablesfalse
 \fi
 \ifnum\NextItem=\Figure
   \FindItem{\Figure}{\NextFigure}
   \ifnum\StackPointer=\m@ne \global\MoreFiguresfalse
   \else
    \GetItemSPAN{\StackPointer}
    \ifnum\ItemSPAN=\Single \def\Zone{\InZoneB}\relax
     \ifFigInZoneC \global\MoreFiguresfalse\fi
    \else
     \def\Zone{\InZoneA}
     \ifFigInZoneB \def\Zone{\InZoneC}\fi
    \fi
   \fi
   \ifMoreFigures\Print{}\FigureItems\fi
 \fi
\ifnum\NextItem=\Table
   \FindItem{\Table}{\NextTable}
   \ifnum\StackPointer=\m@ne \global\MoreTablesfalse
   \else
    \GetItemSPAN{\StackPointer}
    \ifnum\ItemSPAN=\Single\relax
     \ifTabInZoneC \global\MoreTablesfalse\fi
    \else
     \def\Zone{\InZoneA}
     \ifTabInZoneB \def\Zone{\InZoneC}\fi
    \fi
   \fi
   \ifMoreTables\Print{}\TableItems\fi
 \fi
   \MakePageInCompletefalse 
   \ifMoreFigures\MakePageInCompletetrue\fi
   \ifMoreTables\MakePageInCompletetrue\fi
 \repeat
 \ifZoneAFullPage
  \global\TextSize=\z@
  \global\ZoneBSize=\z@
  \global\vsize=\z@\relax
  \global\topskip=\z@\relax
  \vbox to \z@{\vss}
  \eject
 \else
 \global\advance\ZoneBSize by -\ZoneBAdjust
 \global\vsize=\ZoneBSize
 \global\hsize=\ColumnWidth
 \global\ZoneBAdjust=\z@
 \ifdim\TextSize<23pt
 \Warn{}
 \Warn{* Making column fall short: TextSize=\the\TextSize *}
 \vskip-\lastskip\eject\fi
 \fi
}

\def\MakeRightCol{
 \global\TextSize=\ZoneBSize
 \MakePageInCompletetrue
 \MoreFigurestrue
 \MoreTablestrue
 \global\FirstSingleItemtrue
 \global\setbox\ZoneBBOX=\box\VOIDBOX
 \def\Zone{\InZoneB}
 \loop
  \ifMakePageInComplete
 \FindNext
 \ifnum\StackPointer=\m@ne
  \NextItem=\m@ne
  \MoreFiguresfalse
  \MoreTablesfalse
 \fi
 \ifnum\NextItem=\Figure
   \FindItem{\Figure}{\NextFigure}
   \ifnum\StackPointer=\m@ne \MoreFiguresfalse
   \else
    \GetItemSPAN{\StackPointer}
    \ifnum\ItemSPAN=\Double\relax
     \MoreFiguresfalse\fi
   \fi
   \ifMoreFigures\Print{}\FigureItems\fi
 \fi
 \ifnum\NextItem=\Table
   \FindItem{\Table}{\NextTable}
   \ifnum\StackPointer=\m@ne \MoreTablesfalse
   \else
    \GetItemSPAN{\StackPointer}
    \ifnum\ItemSPAN=\Double\relax
     \MoreTablesfalse\fi
   \fi
   \ifMoreTables\Print{}\TableItems\fi
 \fi
   \MakePageInCompletefalse 
   \ifMoreFigures\MakePageInCompletetrue\fi
   \ifMoreTables\MakePageInCompletetrue\fi
 \repeat
 \ifZoneAFullPage
  \global\TextSize=\z@
  \global\ZoneBSize=\z@
  \global\vsize=\z@\relax
  \global\topskip=\z@\relax
  \vbox to \z@{\vss}
  \eject
 \else
 \global\vsize=\ZoneBSize
 \global\hsize=\ColumnWidth
 \ifdim\TextSize<23pt
 \Warn{}
 \Warn{* Making column fall short: TextSize=\the\TextSize *}
 \vskip-\lastskip\eject\fi
\fi
}

\def\FigureItems{
 \Print{Considering...}
 \ShowItem{\StackPointer}
 \GetItemBOX{\StackPointer} 
 \GetItemSPAN{\StackPointer}
  \CheckFitInZone 
  \ifnum\ItemFits=\Yes
   \ifnum\ItemSPAN=\Single
     \ChangeStatus{\StackPointer}{\InZoneB} 
     \global\FigInZoneBtrue
     \ifFirstSingleItem
      \hbox{}\vskip-\BodgeHeight
     \global\advance\ItemSIZE by \TextLeading
     \fi
     \unvbox\ItemBOX\ItemSep
     \global\FirstSingleItemfalse
     \global\advance\TextSize by -\ItemSIZE
     \global\advance\TextSize by -\TextLeading
   \else
    \ifFirstZoneA
     \global\advance\ItemSIZE by \TextLeading
     \global\FirstZoneAfalse\fi
    \global\advance\TextSize by -\ItemSIZE
    \global\advance\TextSize by -\TextLeading
    \global\advance\ZoneBSize by -\ItemSIZE
    \global\advance\ZoneBSize by -\TextLeading
    \ifFigInZoneB\relax
     \else
     \ifdim\TextSize<3\TextLeading
     \global\ZoneAFullPagetrue
     \fi
    \fi
    \ChangeStatus{\StackPointer}{\Zone}
    \ifnum\Zone=\InZoneC \global\FigInZoneCtrue\fi
  \fi
   \Print{TextSize=\the\TextSize}
   \Print{ZoneBSize=\the\ZoneBSize}
  \global\advance\NextFigure \@ne
   \Print{This figure has been placed.}
  \else
   \Print{No space available for this figure...holding over.}
   \Print{}
   \global\MoreFiguresfalse
  \fi
}

\def\TableItems{
 \Print{Considering...}
 \ShowItem{\StackPointer}
 \GetItemBOX{\StackPointer} 
 \GetItemSPAN{\StackPointer}
  \CheckFitInZone 
  \ifnum\ItemFits=\Yes
   \ifnum\ItemSPAN=\Single
    \ChangeStatus{\StackPointer}{\InZoneB}
     \global\TabInZoneBtrue
     \ifFirstSingleItem
      \hbox{}\vskip-\BodgeHeight
     \global\advance\ItemSIZE by \TextLeading
     \fi
     \unvbox\ItemBOX\ItemSep
     \global\FirstSingleItemfalse
     \global\advance\TextSize by -\ItemSIZE
     \global\advance\TextSize by -\TextLeading
   \else
    \ifFirstZoneA
    \global\advance\ItemSIZE by \TextLeading
    \global\FirstZoneAfalse\fi
    \global\advance\TextSize by -\ItemSIZE
    \global\advance\TextSize by -\TextLeading
    \global\advance\ZoneBSize by -\ItemSIZE
    \global\advance\ZoneBSize by -\TextLeading
    \ifFigInZoneB\relax
     \else
     \ifdim\TextSize<3\TextLeading
     \global\ZoneAFullPagetrue
     \fi
    \fi
    \ChangeStatus{\StackPointer}{\Zone}
    \ifnum\Zone=\InZoneC \global\TabInZoneCtrue\fi
   \fi
  \global\advance\NextTable \@ne
   \Print{This table has been placed.}
  \else
  \Print{No space available for this table...holding over.}
   \Print{}
   \global\MoreTablesfalse
  \fi
}


\def\CheckFitInZone{%
{\advance\TextSize by -\ItemSIZE
 \advance\TextSize by -\TextLeading
 \ifFirstSingleItem
  \advance\TextSize by \TextLeading
 \fi
 \ifnum\Zone=\InZoneA\relax
  \else \advance\TextSize by -\ZoneBAdjust
 \fi
 \ifdim\TextSize<3\TextLeading \global\ItemFits=\No
 \else \global\ItemFits=\Yes\fi}
}

\def\BeginOpening{%
  \thispagestyle{titlepage}%
  \global\setbox\ItemBOX=\vbox\bgroup%
    \hsize=\PageWidth%
    \hrule height \z@
    \ifsinglecol\vskip 6pt\fi 
}

\let\begintopmatter=\BeginOpening  

\def\EndOpening{%
  \One
  \egroup
  \ifsinglecol
    \box\ItemBOX%
    \vskip\TextLeading plus 2\TextLeading
    \@noafterindent
  \else
    \ItemNUMBER=\z@%
    \ItemTYPE=\Figure
    \ItemSPAN=\Double
    \ItemSTATUS=\InStack
    \JoinStack
  \fi
}


\newif\if@here  \@herefalse

\def\no@float{\global\@heretrue}
\let\nofloat=\relax 

\def\beginfigure{%
  \@ifstar{\global\@dfloattrue \@bfigure}{\global\@dfloatfalse \@bfigure}%
}

\def\@bfigure#1{%
  \par
  \if@dfloat
    \ItemSPAN=\Double
    \TEMPDIMEN=\PageWidth
  \else
    \ItemSPAN=\Single
    \TEMPDIMEN=\ColumnWidth
  \fi
  \ifsinglecol
    \TEMPDIMEN=\PageWidth
  \else
    \ItemSTATUS=\InStack
    \ItemNUMBER=#1%
    \ItemTYPE=\Figure
  \fi
  \bgroup
    \hsize=\TEMPDIMEN
    \global\setbox\ItemBOX=\vbox\bgroup
      \eightpoint\nostb@ls{10pt}%
      \let\caption=\fig@caption
      \ifsinglecol \let\nofloat=\no@float\fi
}

\def\fig@caption#1{%
  \vskip 5.5pt plus 6pt%
  \bgroup 
    \eightpoint\nostb@ls{10pt}%
    \setbox\TEMPBOX=\hbox{#1}%
    \ifdim\wd\TEMPBOX>\TEMPDIMEN
      \noindent \unhbox\TEMPBOX\par
    \else
      \hbox to \hsize{\hfil\unhbox\TEMPBOX\hfil}%
    \fi
  \egroup
}

\def\endfigure{%
  \par\egroup 
  \egroup
  \ifsinglecol
    \if@here \midinsert\global\@herefalse\else \topinsert\fi
      \unvbox\ItemBOX
    \endinsert
  \else
    \JoinStack
    \Print{Processing source for figure \the\ItemNUMBER}%
  \fi
}


\newbox\tab@cap@box
\def\tab@caption#1{\global\setbox\tab@cap@box=\hbox{#1\par}}

\newtoks\tab@txt@toks
\long\def\tab@txt#1{\global\tab@txt@toks={#1}\global\table@txttrue}

\newif\iftable@txt  \table@txtfalse
\newif\if@dfloat    \@dfloatfalse

\def\begintable{%
  \@ifstar{\global\@dfloattrue \@btable}{\global\@dfloatfalse \@btable}%
}

\def\@btable#1{%
  \par
  \if@dfloat
    \ItemSPAN=\Double
    \TEMPDIMEN=\PageWidth
  \else
    \ItemSPAN=\Single
    \TEMPDIMEN=\ColumnWidth
  \fi
  \ifsinglecol
    \TEMPDIMEN=\PageWidth
  \else
    \ItemSTATUS=\InStack
    \ItemNUMBER=#1%
    \ItemTYPE=\Table
  \fi
  \bgroup
    \eightpoint\nostb@ls{10pt}%
    \global\setbox\ItemBOX=\vbox\bgroup
      \let\caption=\tab@caption
      \let\tabletext=\tab@txt
      \ifsinglecol \let\nofloat=\no@float\fi
}

\def\endtable{%
  \par\egroup 
  \egroup
  \setbox\TEMPBOX=\hbox to \TEMPDIMEN{%
    \hss
    \vbox{%
      \hsize=\wd\ItemBOX
      \ifvoid\tab@cap@box
      \else
        \noindent\unhbox\tab@cap@box
        \vskip 5.5pt plus 6pt%
      \fi
      \box\ItemBOX
      \iftable@txt
        \vskip 10pt%
        \eightpoint\nostb@ls{10pt}%
        \noindent\the\tab@txt@toks
        \global\table@txtfalse
      \fi
    }%
    \hss
  }%
  \ifsinglecol
    \if@here \midinsert\global\@herefalse\else \topinsert\fi
      \box\TEMPBOX
    \endinsert
  \else
    \global\setbox\ItemBOX=\box\TEMPBOX
    \JoinStack
    \Print{Processing source for table \the\ItemNUMBER}%
  \fi
}

\def\UnloadZoneA{%
\FirstZoneAtrue
 \Iteration=\z@
  \loop
   \ifnum\Iteration<\LengthOfStack
    \GetItemSTATUS{\Iteration}
    \ifnum\ItemSTATUS=\InZoneA
     \GetItemBOX{\Iteration}
     \ifFirstZoneA \vbox to \BodgeHeight{\vfil}%
     \FirstZoneAfalse\fi
     \unvbox\ItemBOX\ItemSep
     \LeaveStack{\Iteration}
     \else
     \advance\Iteration \@ne
   \fi
 \repeat
}

\def\UnloadZoneC{%
\Iteration=\z@
  \loop
   \ifnum\Iteration<\LengthOfStack
    \GetItemSTATUS{\Iteration}
    \ifnum\ItemSTATUS=\InZoneC
     \GetItemBOX{\Iteration}
     \ItemSep\unvbox\ItemBOX
     \LeaveStack{\Iteration}
     \else
     \advance\Iteration \@ne
   \fi
 \repeat
}


\def\ShowItem#1{
  {\GetItemAll{#1}
  \Print{\the#1:
  {TYPE=\ifnum\ItemTYPE=\Figure Figure\else Table\fi}
  {NUMBER=\the\ItemNUMBER}
  {SPAN=\ifnum\ItemSPAN=\Single Single\else Double\fi}
  {SIZE=\the\ItemSIZE}}}
}

\def\ShowStack{%
 \Print{}
 \Print{LengthOfStack = \the\LengthOfStack}
 \ifnum\LengthOfStack=\z@ \Print{Stack is empty}\fi
 \Iteration=\z@
 \loop
 \ifnum\Iteration<\LengthOfStack
  \ShowItem{\Iteration}
  \advance\Iteration \@ne
 \repeat
}

\def\B#1#2{%
\hbox{\vrule\kern-0.4pt\vbox to #2{%
\hrule width #1\vfill\hrule}\kern-0.4pt\vrule}
}


\newif\ifsinglecol   \singlecolfalse

\def\onecolumn{%
  \global\output={\singlecoloutput}%
  \global\hsize=\PageWidth
  \global\vsize=\PageHeight
  \global\ColumnWidth=\hsize
  \global\TextLeading=12pt
  \global\Leading=12
  \global\singlecoltrue
  \global\let\onecolumn=\relax
  \global\let\footnote=\sing@footnote
  \global\let\vfootnote=\sing@vfootnote
  \ninepoint 
  \message{(Single column)}%
}

\def\singlecoloutput{%
  \shipout\vbox{\PageHead\pagebody\PageFoot}%
  \advancepageno
  \ifplate@page
    \shipout\vbox{%
      \sp@pagetrue
      \def\sp@type{plate}%
      \global\plate@pagefalse
      \PageHead\vbox to \PageHeight{\unvbox\plt@box\vfil}\PageFoot%
    }%
    \message{[plate]}%
    \advancepageno
  \fi
  \ifnum\outputpenalty>-\@MM \else\dosupereject\fi%
}

\def\ItemSep{\vskip\ItemSepamount\relax}

\def\ItemSepbreak{\par\ifdim\lastskip<\ItemSepamount
  \removelastskip\penalty-200\ItemSep\fi%
}


\let\@@endinsert=\endinsert 

\def\endinsert{\egroup 
  \if@mid \dimen@\ht\z@ \advance\dimen@\dp\z@ \advance\dimen@12\p@
    \advance\dimen@\pagetotal \advance\dimen@-\pageshrink
    \ifdim\dimen@>\pagegoal\@midfalse\p@gefalse\fi\fi
  \if@mid \ItemSep\box\z@\ItemSepbreak
  \else\insert\topins{\penalty100 
    \splittopskip\z@skip
    \splitmaxdepth\maxdimen \floatingpenalty\z@
    \ifp@ge \dimen@\dp\z@
    \vbox to\vsize{\unvbox\z@\kern-\dimen@}
    \else \box\z@\nobreak\ItemSep\fi}\fi\endgroup%
}


\def\gobbleone#1{}
\def\gobbletwo#1#2{}
\let\footnote=\gobbletwo 
\let\vfootnote=\gobbleone

\def\sing@footnote#1{\let\@sf\empty 
  \ifhmode\edef\@sf{\spacefactor\the\spacefactor}\/\fi
  \hbox{$^{\hbox{\eightpoint #1}}$}\@sf\sing@vfootnote{#1}%
}

\def\sing@vfootnote#1{\insert\footins\bgroup\eightpoint\b@ls{9pt}%
  \interlinepenalty\interfootnotelinepenalty
  \splittopskip\ht\strutbox 
  \splitmaxdepth\dp\strutbox \floatingpenalty\@MM
  \leftskip\z@skip \rightskip\z@skip \spaceskip\z@skip \xspaceskip\z@skip
  \noindent $^{\scriptstyle\hbox{#1}}$\hskip 4pt%
    \footstrut\futurelet\next\fo@t%
}

\def\footnoterule{\kern-3\p@ \hrule height \z@ \kern 3\p@}

\skip\footins=19.5pt plus 12pt minus 1pt
\count\footins=1000
\dimen\footins=\maxdimen


\def\landscape{%
  \global\TEMPDIMEN=\PageWidth
  \global\PageWidth=\PageHeight
  \global\PageHeight=\TEMPDIMEN
  \global\let\landscape=\relax
  \onecolumn
  \message{(landscape)}%
  \raggedbottom
}


\output{%
  \ifLeftCOL
    \global\setbox\LeftBOX=\vbox to \ZoneBSize{\box255\unvbox\ZoneBBOX}%
    \global\LeftCOLfalse
    \MakeRightCol
  \else
    \setbox\RightBOX=\vbox to \ZoneBSize{\box255\unvbox\ZoneBBOX}%
    \setbox\MidBOX=\hbox{\box\LeftBOX\hskip\ColumnGap\box\RightBOX}%
    \setbox\PageBOX=\vbox to \PageHeight{%
      \UnloadZoneA\box\MidBOX\UnloadZoneC}%
    \shipout\vbox{\PageHead\box\PageBOX\PageFoot}%
    \advancepageno
    \ifplate@page
      \shipout\vbox{%
        \sp@pagetrue
        \def\sp@type{plate}%
        \global\plate@pagefalse
        \PageHead\vbox to \PageHeight{\unvbox\plt@box\vfil}\PageFoot%
      }%
      \message{[plate]}%
      \advancepageno
    \fi
    \global\LeftCOLtrue
    \CleanStack
    \MakePage
  \fi
}


\Warn{\start@mess}


\catcode `\@=12 



\epsfverbosetrue

\def\numax{{\nu_{\rm max}}}

\def\FF{{\cal F}}
\def\GG{{\cal G}}
\def\CK{{\cal K}}
\def\TT{{\cal T}}

\def\etal{{et al.}}
\def\eg{{e.g.}}
\def\spose#1{\hbox to 0pt{#1\hss}}
\def\Upctil{\spose{\raise 0.8ex\hbox{\hskip4pt$\widetilde{}$}}}
\def\Loctil{\spose{\hbox{\hskip2pt$\widetilde{}$}}}
\def\clm{{\Loctil c}_l^{\,m}}
\def\cnl{{c}^{nl}}
\def\dgr{{^\circ}}

\def\Rsun{R}
\def\Rreal{{\rm I\kern-.2em R}}
\def\d{{\rm d}}

\def\gwig{{\leavevmode\kern0.3em\raise.3ex\hbox{$>$}
\kern-0.8em\lower.7ex \hbox{$\sim$}\kern0.3em}}
\def\lwig{{\leavevmode\kern0.3em\raise.3ex\hbox{$<$}
\kern-0.8em\lower.7ex \hbox{$\sim$}\kern0.3em}}

\hyphenation{Pij-pers}

%
\newcount\apeqnu
\apeqnu=1
\newcount\eqnumber
\eqnumber=1
\def\neqn{{\rm(\the\eqnumber)}\global\advance\eqnumber by 1}
\def\apneqn{{\rm(A\the\apeqnu)}\global\advance\apeqnu by 1}
\def\refeq#1){\advance\eqnumber by -#1 {\rm(\the\eqnumber)} \advance
\eqnumber by #1}
\def\apeqnam#1#2{\immediate\write1{\xdef\
#2{(A\the\apeqnu}}\xdef#1{(A\the\apeqnu}}
\def\eqnam#1#2{\immediate\write1{\xdef\
#2{(\the\eqnumber}}\xdef#1{(\the\eqnumber}}
\newcount\fignumber
\fignumber=1
\def\nfig{\global\advance\fignumber by 1}
\def\refig#1{\advance\fignumber by -#1 \the\fignumber \advance\fignumber by #1}
\def\fignam#1#2{\immediate\write1{\xdef\
#2{\the\fignumber}}\xdef#1{\the\fignumber}}

\def\draft{\headline{\bf File: \jobname\hfill DRAFT\hfill\today}}
\def\ref{\par\noindent
	\hangindent=0.7 true cm
	\hangafter=1}



\Autonumber
\loadboldmathnames

\begintopmatter

\title{A modified $\Rreal^1 \otimes \Rreal^1$ method for helioseismic
rotation inversions}

\author{F.~P.~Pijpers$^{1,2}$ and M.~J.~Thompson$^{3}$}

\smallskip\affiliation{$^1$ Uppsala Astronomical Observatory,
   Box~515, S-751\thinspace{}20 Uppsala, Sweden}
\smallskip\affiliation{$^2$ Present address: Teoretisk Astrofysik Center,
   Danmarks Grundforskningsfond, Institut for Fysik og Astronomi, Aarhus
   Universitet,}
\smallskip\affiliation{\phantom{$^2$} DK-8000 Aarhus C, Denmark}
\smallskip\affiliation{$^3$ Astronomy Unit, Queen Mary and Westfield College,
   University of London, Mile End Road, London E1 4NS, U.K.}

\shortauthor{F.P. Pijpers and M.J. Thompson}

\acceptedline{Accepted . Received }

\abstract{We present an efficient method for two dimensional inversions
for the solar rotation rate using the Subtractive Optimally Localized
Averages (SOLA) method and a modification of the $\Rreal^1\otimes
\Rreal^1$ technique proposed by Sekii (1993a,b). The SOLA method is
based on explicit construction of averaging kernels similar to the
Backus-Gilbert method. The versatility and reliability of the SOLA
method in reproducing a target form for the averaging kernel, in
combination with the idea of the $\Rreal^1 \otimes \Rreal^1$ decomposition,
results in a computationally very efficient inversion algorithm.
This is particularly important for full 2-D inversions of helioseismic
data in which the number of modes runs into at least tens of thousands.
}

\keywords{ Sun : oscillations of -- Sun : rotation of --
             Sun : structure of -- Numerical methods }

\maketitle

\section{Introduction}

The solar 5-minute oscillations can be described as a superposition
of eigenmodes of non-radial pulsation. Each mode is identified
by three integers $(n,l,m)$,
where $l$ and $m$ are the degree
and order respectively of a spherical harmonic,
and $n$ is essentially the number of radial
nodes in the displacement eigenfunction: $m$ can take all values from $-l$
to $+l$.
In a spherically symmetric, nonrotating star,
the frequency of an eigenmode would be independent of $m$ and thus there
would be
multiplets of $(2l+1)$ modes with identical frequencies, each multiplet
corresponding to an $(n,l)$ pair.
Rotation lifts this
($2l+1$)-fold degeneracy. The difference in frequency between modes in the same
multiplet is called the (rotational)
frequency splitting. The frequency splitting is
determined by the rotation rate inside the Sun and can be used in an
inverse problem to probe the Sun's internal rotation. In particular, they
enable one to perform 2-D inversions for the rotation rate as a function of
radius and latitude.

Large helioseismic data sets should soon be available from various
observational campaigns, notably the Global Oscillations Network
Group (GONG) and MDI-SOI on board the SOHO satellite.
To make optimal use of these data,
the algorithms for full 2-D helioseismic inversions need
to become efficient in handling several hundreds of thousands, or even
millions, of data (modes) simultaneously. Optimally localized averages
(OLA) techniques, which have proved very popular for 1-D helioseimic
inversions involving only a few thousand data, require a matrix to be
inverted whose order is the total number of data. This is prohibitively
expensive computationally in the 2-D case -- though one may be able to
make the computation tractable by preprocessing to reduce the number of
data to which the OLA is applied (Christensen-Dalsgaard {\etal} 1994).
Least-squares techniques ({\eg} Schou {\etal} 1994),
which require a matrix to be inverted whose
order is the number of base functions, are also expensive in the 2-D
case, where one might want a discretization of {\eg} 200 bins in radius and
100 in latitude which gives a matrix of order $2\times 10^4$.

Sekii (1993a,b) has exploited the fact that the rotation kernels are
nearly separable in radius $r$ and colatitude $\theta$ to develop a
so-called $\Rreal^1 \otimes \Rreal^1$ inversion technique in which the
true kernels are approximated by ones which are exactly separable. This
results in a problem where the order of the largest matrix one has to
invert is only the number of $(n,l)$-multiplets, which is only a few thousand.

Here we propose a modification to Sekii's $\Rreal^1 \otimes \Rreal^1$ method,
in which the
small deviations from separability of the kernels are taken into account.
The computational burden is the same as for Sekii's approach; hence it is
just as efficient. Again the problem is reduced to a series of 1D
inversions, for which we use the subtractive optimally localized averages
(SOLA) method of Pijpers \& Thompson (1992, 1994; hereafter PT1, PT2). The
SOLA has the advantage over other 1D inversion methods that it is
possibly to keep close control over the averaging kernels that it produces:
this turns out to be important for the $\Rreal^1 \otimes \Rreal^1$
inversion, as we shall see below (cf. also Sekii 1994).

\section{The $\Rreal^1 \otimes \Rreal^1$ method}

In a spherically symmetric, nonrotating star, the frequency
$\omega_{nl}$ of a
spheroidal mode of oscillation of radial order $n$ and degree $l$ is
independent of the mode's azimuthal order $m$, and the displacement
eigenfunction is
$$
\left( \xi_{nl}(r), L^{-1}\eta_{nl}{\partial\over\partial\theta},
{L^{-1}\eta_{nl}\over\sin\theta}{\partial\over\partial\phi}\right)
P_l^m(\cos\theta)e^{im\phi}
\eqno\neqn
$$
with respect to spherical polar coordinates $(r,\theta,\phi)$: here
$\xi_{nl}$ and $\eta_{nl}$ are calculable functions, given a solar model;
$L = {\sqrt{l(l+1)}}$; and $P_l^m$ is the Legendre function of
degree $l$ and order $m$.

For a slowly rotating star the frequencies also depend on $m$. According
to linear perturbation theory, which is an excellent approximation for the
Sun, the difference between frequencies of modes with the same values
of $n$ and $l$, but opposite $m$, is given in terms of the
eigenfunctions of the nonrotating star by
$$
\omega_{nlm} - \omega_{nl\,-m}\ = 2m D_{nlm}
$$
where
\eqnam{\RotRate}{RotRate}
$$
D_{nlm}\ =\
=\ \int\limits_{-1}^1 \int\limits_0^1  K_{nlm}(r,\theta)
\Omega(r,\theta) \d r\,\d\cos\theta\,.
\eqno\neqn
$$
Here and in the following,
we have made the radial variable $r$ dimensionless by dividing
it by the surface radius.

An OLA inversion for the rotation amounts to finding coefficients
$\{c_{nlm}(r_0,\theta_0)\}$ so that
$$
\eqalign{
{\bar\Omega} (r_0,\theta_0) \equiv \sum_{nlm} c_{nlm}&(r_0,\theta_0)
D_{nlm} =\cr &\int\int \sum c_{nlm} K_{nlm} \Omega \,\d r\,\d\cos\theta \,.\cr}
\eqno\neqn
$$
is a localized average of the actual rotation rate $\Omega$ near
$r = r_0, \theta=\theta_0$.
To find the coefficients with a naive application of OLA
would require a matrix to be inverted whose size was
the total number of
observed eigenmodes i.e. all avaliable $(n,l,m)$ combinations. This is
prohibitively expensive with a very large mode set.
This paper
is concerned with finding suitable coefficients in a computationally less
expensive way, by exploiting properties of the kernels $K_{nlm}$.

The $K_{nlm}$ are given by
\eqnam{\Separable}{Separable}
$$
K_{nlm} (r, \theta)\ =\ F_1^{nl}(r) G_1^{lm}(\theta)\;+\;
F_2^{nl}(r) G_2^{lm}(\theta)\,,
\eqno\neqn
$$
where
\eqnam{\Fodef}{Fodef}
$$
F^{nl}_1\ \equiv\ \rho(r) r^2 \left[ \xi_{nl}^2(r) - 2L^{-1} \xi_{nl}(r)
\eta_{nl} (r) + \eta_{nl}^2(r) \right] / I_{nl}
\eqno\neqn
$$
\eqnam{\Ftdef}{Ftdef}
$$
F^{nl}_2\ \equiv\ \rho(r) r^2 \left[ \eta_{nl}^2(r) \right]
/ I_{nl}\,,
\eqno\neqn
$$
$\rho$ being the density and and $\xi_{nl}$, $\eta_{nl}$ the components of
the displacement eigenfunction in the nonrotating star, and
$$
I_{nl}\ =\ \int\limits_0^1 \d r\, \rho r^2 (\xi_{nl}^2 +
\eta_{nl}^2 )\,;
\eqno\neqn
$$
and
\eqnam{\Godef}{Godef}
$$
G^{lm}_1\ \equiv\ { (l - \vert m \vert)! \over (l + \vert m \vert)!} (l+1/2)
\left[ P_l^m(u) \right]^2
\eqno\neqn
$$
\eqnam{\Gfunrel}{Gfunrel}
$$
G^{lm}_2\ \equiv\ {1\over 2} L^{-2} (1-u^2) {\d^2
G^{lm}_1\over\d u^2}
\eqno\neqn
$$
$$
u\ \equiv\ \cos\theta\;.
$$
Note that our definition of $\eta_{nl}$ differs from that used by Sekii
(1993a) by a factor of $L$. The ratio of the amplitudes of
our $\xi_{nl}$ and $\eta_{nl}$ is roughly the ratio of the vertical
wavenumber to the horizontal wavenumber (since for p modes the waves
are almost longitudinal): thus for p modes, which propagate more nearly
vertically than horizontally except near their turning points,
$\xi_{nl}$ is larger in magnitude than $\eta_{nl}$.
More quantitatively, using the simplest asymptotics,
the ratio of the amplitudes of $F_1^{nl}$ to
$F_2^{nl}$ is of order $\omega^2 r^2 / L^2 c_S^2$ in the acoustic cavity of
the mode, $\omega$ being the mode's frequency and $c_S$ the local
adiabatic sound speed. Although this is of order unity near the lower
turning point, it is much larger than unity elsewhere in the acoustic
cavity. The angular functions $G_1^{lm}$, $G_2^{lm}$ are of similar
magnitude, since the two derivatives of $G_1^{lm}$ each produce a
factor of $L$ and these cancel the factor of $L^{-2}$.
Hence, as Sekii (1993a,b) observed, the contribution from
$F_2^{nl} G_2^{lm}$ is generally small compared to that from
$F_1^{nl} G_1^{lm}$, and so the original problem can be approximated by
the separable problem
$$
D_{nlm}\ \equiv {\omega_{nlm} - \omega_{nl\,-m} \over 2 m}\approx
\int\int F_1^{nl}(r) G_1^{lm}(u)\,\d r\d u\,.
\eqno\neqn
$$
A further aspect is that
$G_1^{lm}$ is everywhere positive, whereas $G_2^{lm}$ oscillates to both
positive and negative values -- increasingly so for larger values
of  $(l-|m|)$ -- and this too helps make the integrated contribution from the
$F_2^{nl} G_2^{lm}$ small in general.

The essence of Sekii's idea for exploiting the separability is as follows.
(We shall work within the framework of SOLA, which is the 1D method we
shall employ.) Firstly, for each $l$ one seeks coefficients
$\clm(\theta_0)$ such that
\eqnam{\Simple}{Simple}
$$
\sum_{m=1}^{l} \clm (\theta_0) G_1^{lm} (\theta)
\ \approx\ {\Upctil T}_l(u - u_0)\,,
\eqno\neqn
$$
where $u_0 = \cos\theta_0$, and
${\Upctil T}_l(u - u_0)$ is some chosen target form that is
peaked around $\theta = \theta_0$ and small elsewhere. (The means by which
such coefficients may be sought is described in PT1, PT2.) Then
\eqnam{\latComb}{latComb}
$$
\eqalign{
\sum_{m=1}^{l} \clm &(\theta_0) D_{nlm} =\cr
&= \int \kern-.35em \int F_1^{nl}\left(
\sum_{m=1}^{l} \clm(\theta_0) G_1^{lm} (\theta)\right)
\Omega(r,\theta)\,\d r\d u\cr
&\equiv\ \int F_1^{nl}(r) \langle\Omega\rangle_l^{(\theta_0)}(r)\,\d r\,,
\cr}
\eqno\neqn
$$
where $\langle\Omega\rangle_l^{(\theta_0)}\approx
\int {\Upctil T}_l( u - u_0) \Omega(r,\theta)\,\d u$. The
second and final step is then to choose further coefficients
$\cnl(r_0)$ such that
\eqnam{\RadKer}{RadKer}
$$
\sum_{nl} \cnl(r_0) F_1^{nl}(r)\ \approx\ T(r-r_0)
\eqno\neqn
$$
where similarly $T(r-r_0)$ is a chosen target function that is localized
about some radius, $r_0$. Now if ${\Upctil T}_l(u - u_0)$ were
in fact independent of $l$, so
$\langle\Omega\rangle_l = \langle\Omega\rangle$ say,
eqs. \latComb) and \RadKer) would imply that
$$
\eqalign{
\sum_{nlm}\cnl {(r_0)}\clm(\theta_0) D_{nlm}
&\approx\int\sum_{nl}\cnl{(r_0)} F_1^{nl}(r)\,\times\cr
&\hskip 2.3cm
\langle\Omega\rangle^{(\theta_0)}(r)\,\d r\cr
&\ \equiv\langle\langle\Omega\rangle\rangle^{(r_0,\theta_0)}\,,\cr}
\eqno\neqn
$$
where
$\langle\langle\Omega\rangle\rangle^{(r_0,\theta_0)} \approx
\int\int T(r-r_0) {\Upctil T}(u -u_0) \Omega(r,\theta)\,\d r
\d \theta$. This then completes the $\Rreal^1 \otimes \Rreal^1$ inversion.

There are some drawbacks to the simple
$\Rreal^1 \otimes \Rreal^1$ procedure outlined above. One is that the
linear combinations on the left-hand side of \Simple) need to be essentially
independent of $l$. A practical matter is that
choosing the angular target functions to be independent
of $l$ does not guarantee that $\sum \clm {(\theta_0)}
G_1^{lm}(\theta)$
will itself be independent of the degree. However, SOLA is better than other
commonly-used methods in forcing the linear combination to accurately
resemble a given form (Dziembowski {\etal} 1994, Sekii 1994).
A second point is that, in making all the target functions independent
of $l$ and $r_0$, the angular resolution implied by \Simple) is the same
at all target radii. But because we have many more $m$ values for the
shallowly penetrating high-degree modes than for the deeply penetrating
modes of low degree, we should be able to achieve much better angular
resolution in the outer layers of the Sun than in its deep interior.
[This point was appreciated by Sekii (1993b); but one of the
contributions of the present paper is to suggest how the angular
resolution might be chosen appropriate to the target depth.]
Thus we need to relax the restriction that the
combinations of latitudinal kernels are independent of both degree and radial
target. In order to preserve the property that the radial inversion can be
treated as a 1-D inversion, we shall allow the target functions
$\TT_l$, and hence
the coefficients $\clm$, to depend on the radial as well as the latitudinal
location of the target point. Thirdly, although it may well be an excellent
approximation to neglect the $F_2^{nl}G_2^{lm}$ terms
for most presently observed modes, we hope that
forthcoming datasets will extend to even lower frequency low-degree modes,
for which Sekii's approximation is less good. For this reason, we wish to
include the $F_2^{nl}G_2^{lm}$ contribution to the kernels while retaining
the advantage of the basic $\Rreal^1 \otimes \Rreal^1$ concept.

We therefore now introduce a modification to Sekii's method to take into
account
the $F_2^{nl} G_2^{lm}$ terms when trying to form a localized average.
Specifically, in the latitudinal localization, we seek coefficients
$\clm {(r_0, \theta_0)}$  to minimize
\eqnam{\Minmjt}{Minmjt}
$$
\eqalign{
\int_{-1}^{1} &\left( \sum_{m=1}^l \clm G_1^{lm}
- \TT_{G_1}(u - u_0)\right)^2 \,\d u\cr
& \;+\;
\beta_l^2 \int_{-1}^{1} \left( \sum_{m=1}^l \clm
G_2^{lm} - \TT_{G_2} (u - u_0) \right)^2 \,\d u\cr
& \;+\;
\mu\sum_{m\,m^\prime} {\bf E}_{m\,m^\prime}^{(l)}
\clm {\Loctil c}_l^{\, m^\prime}  \,.}
\eqno\neqn
$$
Here
$$
\mu\ \equiv\ \mu_0\left( \sum_{m\,m'} {\bf E}_{m\,m'}^{(l)}\right)^{-1}\;,
\eqno\neqn
$$
$\mu_0$ being an adjustable parameter; ${\bf E}_{m\,m^\prime}^{(l)}$ is a
suitable error-covariance matrix (see below).
Minimizing the error term and obtaining a well-localized kernel
are opposing aims ({\eg} CDST) and the error weighting parameters
$\mu_0$ are used to obtain a compromise between them.
We simultaneously
force the linear combinations of $G_1^{lm}$ and of $G_2^{lm}$ to
resemble the chosen target functions
$\TT_{G_1}$ and $\TT_{G_2}$, which depend on $l$, $r_0$ and $\theta_0$;
and $\beta_l$ is an adjustable parameter that weights the
relative importance of matching $\TT_{G_1}$ and matching $\TT_{G_2}$.
It is convenient to define function vectors
$$
{\bf F}^{nl} = \pmatrix{F_1^{nl}\cr F_2^{nl}},\quad
{\bf G}^{lm} = \pmatrix{G_1^{lm}\cr G_2^{lm}} \;,
\eqno\neqn
$$
and
$$
\GG^l\; =\; \sum\limits_{m} \clm {(\theta_0)} {\bf G}^{lm},\quad
{\bf \TT}_G^{l} = \pmatrix{\TT_{G_1}\cr \TT_{G_2}} \;.
\eqno\neqn
$$
Note that we shall generally omit the superscript $l$ on $\GG^l$ and
${\bf \TT}_G^l$, except when we wish to emphasize their dependence
on the degree $l$.
Then expression \Minmjt) can be written in suggestive form
\eqnam{\Minimize}{Minimize}
$$
\eqalign{
\int\limits_{-1}^{1} \d u\, (\GG\;-\;\TT_G)^T
{\pmatrix{1&0\cr 0&\beta_l^2\cr}}&
(\GG\;-\;\TT_G)\cr
&+\ \mu\sum_{m\, m^\prime} {\bf E}_{m\, m^\prime}^{(l)}
\clm \,{\Loctil c}_l^{\, m^\prime}\,.}
\eqno\neqn
$$
In effect, we are seeking to localize a vector of functions
rather than a single function, so that the inversion can once again
be carried out as a sequence of two 1-D inversions but
without the approximation Sekii (1993a,b) employed.

If a clean two-dimensional localization is finally to be achieved, it
is desirable that the components $\TT_{G_1}$ and $\TT_{G_2}$ of
$\TT_G$ should have similar shapes. Thus we choose to make the
second component the same as the first, up to a multiplicative scalar.

For convenience later we also impose an exact constraint on the
first component of $\GG$, that it be normalized such that
\eqnam{\GConstraint}{GConstraint}
$$
\int\limits_{-1}^{1} \d u\ {\GG}_1^l (\theta_0 ,\theta)\ =\ 1 \ .
\eqno\neqn
$$

If the latitudinal localization is successful, then
\eqnam{\Gsumming}{Gsumming}
$$
\eqalign{
\sum_{m=1}^{l} \clm {(r_0,\theta_0)} &D_{nlm}\ =
\int \int {\bf F}^{nl}\cdot\GG
\Omega(r,\theta)\,\d r\d u\cr
&\approx\int\int {\bf F}^{nl}\cdot\TT_G
\Omega(r,\theta)\,\d r\d u\cr
&=\int\int (F_1^{nl} + \zeta^{nl} F_2^{nl})\, \TT_{G_1} \,\Omega \,\d r
\,d u\,,
\cr}
\eqno\neqn
$$
where $\zeta^{nl}$ is essentially the ratio of $\TT_{G_2}$ to $\TT_{G_1}$.

The second and final
step in the two dimensional inversion is to find a second set of
coefficients such that~:
\eqnam{\Fcombine}{Fcombine}
$$
{\FF}(r_0 ,r) \equiv \sum\limits_{nl} c^{nl} (r_0)
\left[ F_1^{nl}(r) + \zeta^{nl} F_2^{nl}(r) \right]
\eqno\neqn
$$
is peaked around $r=r_0$ and is small everywhere else.

Actually the $\zeta^{nl}$ that enters in \Fcombine), which is defined
explicitly below in equation (29), is not precisely the
ratio of the
components of the latitudinal target function we use, though they are
closely related. For our latitudinal localization we use
\eqnam{\Tarfundef}{Tarfundef}
$$
{\TT}_G^l = \pmatrix{
{\displaystyle 1\over\displaystyle f_{\theta}\Delta_{\theta}}
\exp\left[ -\left({\displaystyle u - u_0 \over
\displaystyle \Delta_{\theta}} \right)^2 \right]
\cr
\phantom{nothing}
\cr
{\displaystyle -1 \over\displaystyle f_{\theta} \Delta_{\theta}^3 L^2}
\exp\left[ -\left({\displaystyle u - u_0 \over
\displaystyle \Delta_{\theta}}\right)^2
\right] }\, .
\eqno\neqn
$$
The factor $f_{\theta}$ is a normalization factor
which is included to make the total integral of $\TT_{G1}$ equal to unity.

The radial target function is
$$
{\TT}_F = {1\over f_{r}\Delta_{r}} \exp\left[ - \left({ r-r_0 \over
\Delta_{r} }\right)^2\right]
\eqno\neqn
$$
The expression to minimize is now
$$
\int\limits_0^1 \d r \left( \FF - \TT_F \right)^2 + \mu \sum\limits_{nl n'l'}
{\bf E}_{nl\, n'l'} c^{nl} c^{n'l'}\,.
\eqno\neqn
$$

The free parameters $\Delta_r$, $\Delta_\theta$
should in general be functions of target position.
The choice of $\Delta_r$, $\Delta_\theta$ at different radii
and at different
latitudes is an important issue. We have used the natural scaling that
was presented in PT2 (cf. Thompson 1993) for radial and latitudinal
resolution:
\eqnam\Delteqs{Delteqs}
$$
\eqalign{
\Delta_r (r_0)\ &= {\alpha_r\over 8}\ {c_S(r_0) \over \Rsun \numax}\cr
\Delta_\theta (r_0, \theta_0)\ &= \alpha_{\theta} {\Delta_r \over r_0}
\sqrt{1+\epsilon - u_0^2} \cr
}
\eqno\neqn
$$
where $\alpha_r$ and $\alpha_\theta$ are constants of proportionality,
independent of radius. To avoid problems at $u_0 = 1$ we include the
small number $\epsilon = 0.075$.
The strategy is to optimize the target widths at one
target location and then use the relation \refeq1) to calculate the
target widths for all other locations.
One should note that it is at this point that the dependence
of the angular resolution on radius is introduced. The choice of functional
dependence in \Delteqs) expresses the fact that one expects the attainable
physical horizontal resolution and radial resolution to be very similar
at all radii.

\fignam{\flowchart}{flowchart}
\beginfigure{1}
\epsfxsize=8.5cm
\epsfbox{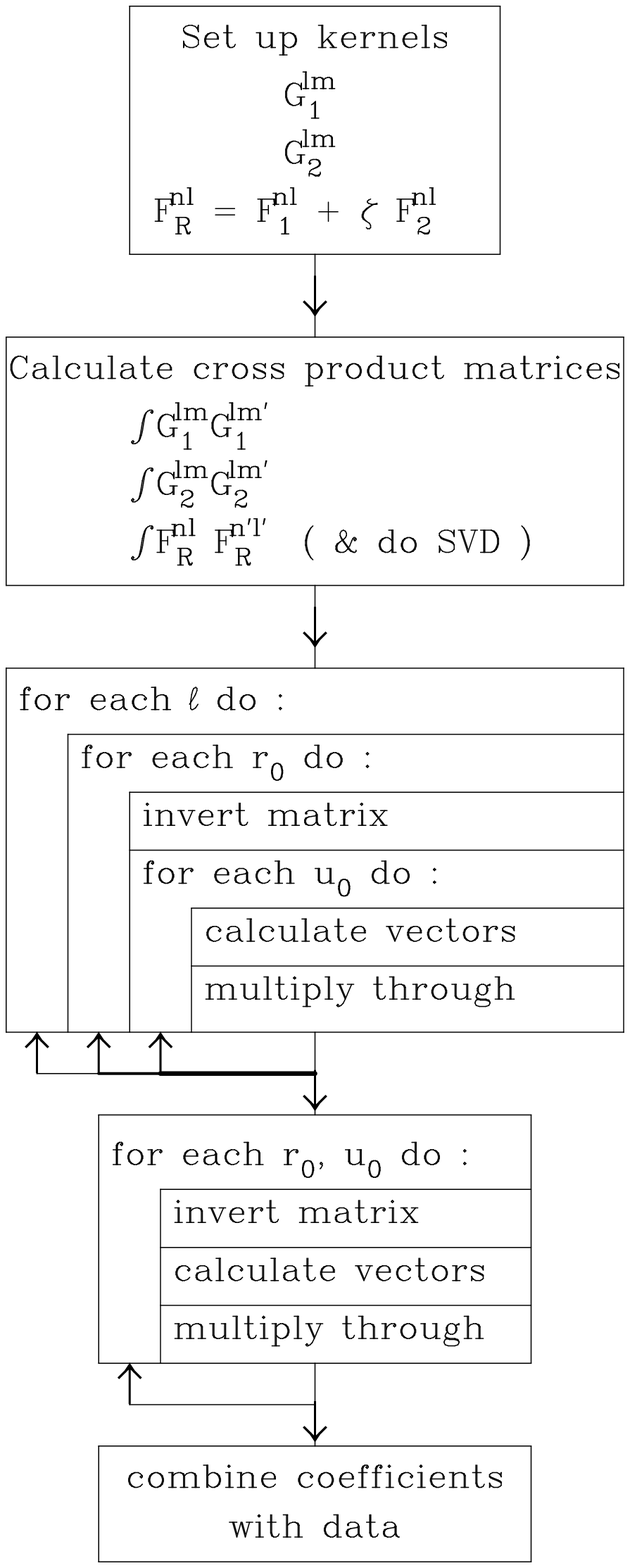}
\caption{{\bf Figure \flowchart} Flowchart showing the principal steps
in the modified $\Rreal^1 \otimes \Rreal^1$ method presented here.}
\endfigure\nfig

The factor $\zeta^{nl}$ that appears in the radial inversions \Fcombine)
is necessary
because the values of the two latitudinal averaging kernels at the
location of their maxima differs due to the relation \Gfunrel). The
factor $\zeta^{nl}$ can compensate for this in the radial inversions so that
in the final reconstructions the $F_2^{nl}$ do not suddenly acquire
a much smaller or larger weight due to the multiplication by the $G_2^{lm}$.
Using \Gfunrel) and \Tarfundef) the ratio of the values of the two
latitudinal kernels at their maximum is approximately~:
\eqnam{\Gfunrat}{Gfunrat}
$$
{\GG_2 (\theta_0, \theta = \theta_0) \over \GG_1(\theta_0, \theta =
\theta_0) }\ =\ - {(1- u_0^2) \over
\Delta_{\theta}^2 L^2}
\eqno\neqn$$
The dependence of $\zeta^{nl}$ on the latitudinal resolution width implies
that a straightforward application of this recipe means that the radial
kernels to be used in the inversion depend on the choice of the target
localization radius. This would defeat the purpose of using SOLA, because
it would require calculating and inverting a new matrix for each new
target position $r_0$. However, one can make use of the fact that any
mode will be used primarily at target radii $r_0$ close to the turning
point of that mode $r_{t}$.
In other words the coefficient for this mode is only
large if the averaging kernel to be constructed peaks near the turning
point of that mode. Using this information and the scaling law for the
latitudinal resolution widths, $\zeta^{nl}$ becomes~:
\eqnam{\GratTP}{GratTP}
$$
\zeta^{nl}\ \propto\ {\GG_2 (\theta_0, \theta = \theta_0) \over \GG_1(\theta_0,
\theta = \theta_0) }\ \sim\ - \left( {4 \over \pi \alpha_\theta \alpha_r}
{\nu_{\rm max} \over
\nu} \right)^2
\eqno\neqn
$$
If the absolute value of this factor is allowed to become much larger
than unity it turns out to over-compensate for the effect of the
amplitude of $\GG_2$ in the radial inversion.
Essentially what we wish to do with $\zeta^{nl}$ is to make sure that
the product $F_2 G_2$ will indeed be treated as a small correction term,
and hence $\zeta^{nl}$ should be less than unity. Thus in practice we
achieve this while retaining the $\nu$-dependence in \GratTP) by
defining
\eqnam\zetainuse{zetainuse}
$$
\zeta^{nl}\ =\ -\,\left({\nu_{\rm min} \over \nu}\right)^2 \,.
\eqno\neqn
$$
Equation \zetainuse) is used for all except the lowest degree modes
for which the turning point is not close to the point at which the mode
is actually used in localized kernels.

The factors $\beta_l$ are used to compensate for the difference in
the absolute value of the integrals of $\TT_{\GG_1}$ and $\TT_{\GG_2}$.
We take $\beta_l = \langle \Delta_{\theta}^2 L^2\rangle$ where the
$\langle\rangle$ denotes a simple average over the latitudinal target
points.

Now combining \Gsumming) with \Fcombine), this time keeping track of
data errors $\epsilon_{nlm}$,  yields~:
\eqnam{\Fsumming}{Fsumming}
$$
\sum\limits_{nl} c^{nl} (r_0) \sum\limits_{m} \clm (r_0, \theta_0)
D_{nlm} \ \approx
$$
$$
\ \approx\
\int\limits_0^1 \d r\, \sum\limits_{nl} c^{nl}
\left[ F_1^{nl}(r) +\zeta^{nl} F_2^{nl}(r) \right]
\langle \Omega ( r, \theta_0)\rangle^l \, +\,
$$
$$
\hskip 4.0truecm +\,\sum\limits_{nl} c^{nl} \sum_m \clm
\epsilon_{nlm}\ .
$$
$$
\ =\ \int\limits_0^1 \d r\, \FF(r_0, r)
\langle \Omega ( r, \theta_0)\rangle \, +\,
\sum\limits_{nl} c^{nl} \sum_m \clm \epsilon_{nlm}\ .
$$
$$
\ =\ \langle \Omega ( r_0, \theta_0)\rangle \, +\,
\sum\limits_{nl} c^{nl} \sum_m \clm \epsilon_{nlm}\ .
\eqno\neqn
$$
The constraint
\eqnam{\FConstraint}{FConstraint}
$$
\int\limits_0^1 \d r\, {\FF} (r_0 ,r)\ =\ 1
\eqno\neqn
$$
is now sufficient to ensure that the complete 2-D averaging kernel that
is constructed is correctly normalized~:
$$
\int\limits_0^1 \d r \int\limits_{-1}^{1} \d u
\ \CK(r_0,\theta_0;r,\theta)\ =\ 1\ .
\eqno\neqn
$$
where
\eqnam{\Avkern}{Avkern}
$$
\CK (r_0, \theta_0; r,\theta) \equiv
\sum\limits_{nl} c^{nl} \sum\limits_{m} \clm
\left[ F^{nl}_1 G^{lm}_1 + F^{nl}_2 G^{lm}_2 \right].
$$
$$
\eqno\neqn
$$
The difference between the target form and the actually
constructed kernel will give rise to a `systematic' error a measure of
which is $\chi$~:
$$
\chi\equiv\int\limits_0^1 \d r \kern-0.35em\int\limits_{-1}^{1}
\d u \left[ \sum\limits_{nl} \sum\limits_{m} c^{nl}
\clm \left( F_1^{nl} G_1^{lm} + F_2^{nl} G_2^{lm} \right)
\right.
$$
$$
\hskip 4truecm
- \TT_F \TT_{G1} \Biggr]^2
\eqno\neqn
$$

The error covariance matrix ${\bf E}_{m\,m^\prime}^{(l)}$ used in
\Minmjt) and \Minimize) is derived from that of the observed
frequencies. Since the localization in colatitude is done at fixed $l$, the
indices $m$ and $m^\prime$
run over the range of $m$, viz from $1$ to $l$. The appropriate
covariance matrix is therefore $E_{n l m, n l m^\prime}$ where $l$ is the
degree under consideration and $n$ can be any order such that the ($n,l$)
multiplet is in the dataset. The overall magnitude of $E$ is irrelevant,
since this is taken out by the scaling in $\mu$. One should note that the
error covariance of $m$ and $m^\prime$ for a given $l$ is taken to be
independent of the value of $n$. This restricts the permitted data error
covariance matrices.
Of course one can proceed even if the errors do
not satisfy this assumption, but the inversion will be less ``optimal''.

\fignam{\resolution}{resolution}
\beginfigure{2}
\epsfxsize=8.5cm
\epsfbox{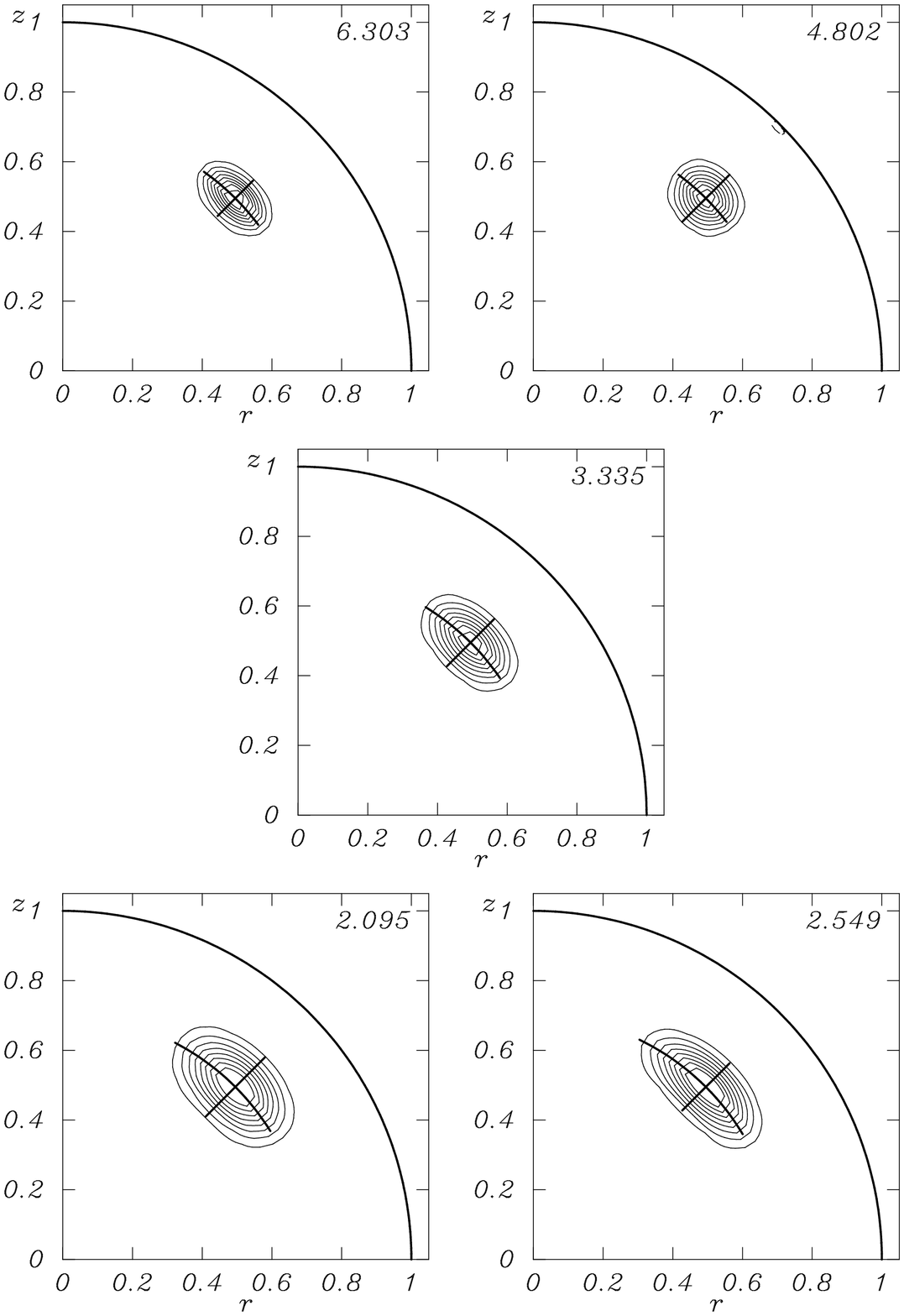}
\caption{{\bf Figure \resolution} Constructed unimodular averaging kernels
for various values of latitudinal and radial resolution widths. Each panel
is labelled with the associated error in the solution, in nHz. Middle panel
$\alpha_r = 8.0$, $\alpha_\theta = 1.5$. Upper left $\alpha_r = 6.0,
\alpha_\theta=1.5$, upper right $\alpha_r = 8.0, \alpha_\theta=1.0$,
lower left $\alpha_r = 10.0, \alpha_\theta = 1.5$, lower right
$\alpha_r = 8.0, \alpha_\theta=2.0$. The target location is $r_0 = 0.7 R,
\theta_0 = 45\dgr$ in all cases. The highest contour is $\sim 10\%$ below
the peak value and the other contours are spaced in $1/8$ of that value.
The crossbar has an extension that corresponds that of a Gaussian at
the lowest contour level which is at $\vert r-r_0 \vert = 1.478 \Delta_r,
\, \vert u - u_0 \vert = 1.478 \Delta_\theta$. All error weighting
parameters $\mu_0 = 0.1$ }
\endfigure\nfig

The error covariance matrix
${\bf E}_{nl\, n'l'}$ in the radial inversion is also
derived from that of the observed frequencies. We assume that
the data error covariance matrix can be factorised, into an error
(co-)variance between multiplets and an error (co-)variance for
$m$ given any fixed $l$. If this is not in fact the case, one can still
approximate the errors in this manner, but once again the inversion will
presumably not be optimal.
In the radial inversions the matrix $E_{nl\, n'l'}$ is obtained by~:
$$
\eqalign{
E_{nl\, n'l'}\ =\ e_{nl\, n'l'} \left(\sum\limits_{m\, m'}
{\bf E}_{m\, m'}^{(l)} \right.&\clm {\Loctil c}_l^{\, m'}
\Biggr)^{1/2} \times \cr
\left(\sum\limits_{m\, m'} \right.&{\bf E}_{m\, m'}^{(l')}
{\Loctil c}_{l'}^{\, m} {\Loctil c}_{l'}^{\, m'} \Biggr)^{1/2}\;. \cr
}
\eqno\neqn
$$

\fignam{\ourtableau}{ourtableau}
\beginfigure*{3}
\epsfxsize=16.5cm
\epsfbox{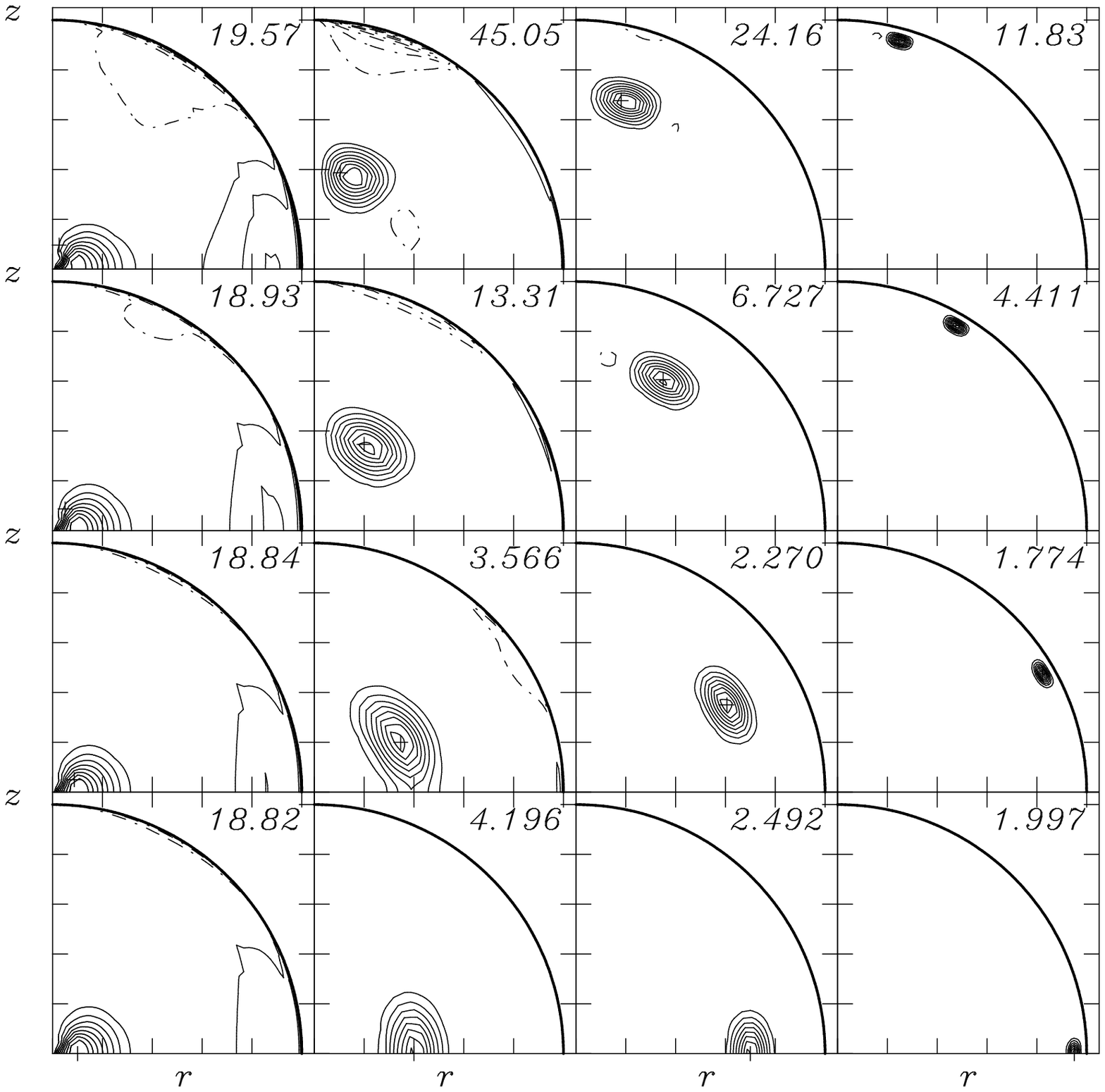}
\caption{{\bf Figure \ourtableau} Unimodular averaging kernels at radii
$r_0 = 0.1, 0.4, 0.7, 0.95$ and co-latitudes $\theta_0 = 15\dgr,
30\dgr , 60\dgr , 90\dgr $.
Target locations are indicated by crosses; standard errors are as quoted.
Localization parameters were $\alpha_r = 8.0$
$\alpha_\theta = 1.5$ and error weighting factor $\mu_0 = 0.1$ for all
kernels.}
\endfigure\nfig

\noindent
Here $e_{nl\, n'l'}$ is a matrix which expresses the error variance/
co-variance for all $(n,l)$-multiplets. We conclude this presentation
of the modified $\Rreal^1 \otimes \Rreal^1$ method by outlining the
principle steps in a flow chart (Fig. {\flowchart}).
The first and second box represent
steps that are preparatory to the actual inversion, the third box
represents the latitudinal part of the inversion and the fourth box the
radial part. The final box represents the process of combining the
linear coefficients from the inversion with the data.

\section{Results}

To illustrate the results which may be obtained with the method, we have
applied it to artificial splitting data. The
modeset is exactly the same as used in the GONG Hare and Hounds exercise
(Gough \& Toomre 1993): in brief, the set consists of f and p modes
below 5mHz with
$$
l=1,2,\dots,16,18,\dots,50,55,\dots,150,160,\dots,250
$$

\noindent
making a total of 69662 individual modes with positive $m$ values
(in 1380 $n,l$ multiplets). The assumed uncertainties on the frequency
differences $\omega_{nlm} - \omega_{0nl}$ were of the form
$\sigma_l(\nu)\equiv f(\nu) g(l)$, the functions $f$ and $g$
being estimated from Fig. 3.5 of Gough \& Toomre (1993), and so are roughly
the same as the assumed uncertainties in the Hare and Hounds exercise. The
range of $\sigma_l(\nu)$ was from about 4nHz at low degree and frequency to
roughly 700nHz at high degree and frequency. Note that under these
assumptions, the uncertainties in the $D_{nlm}$ are $f(\nu_{nl})g(l)/m$.

\fignam{\Taktableau}{Taktableau}
\beginfigure*{4}
\epsfxsize=16.5cm
\epsfbox{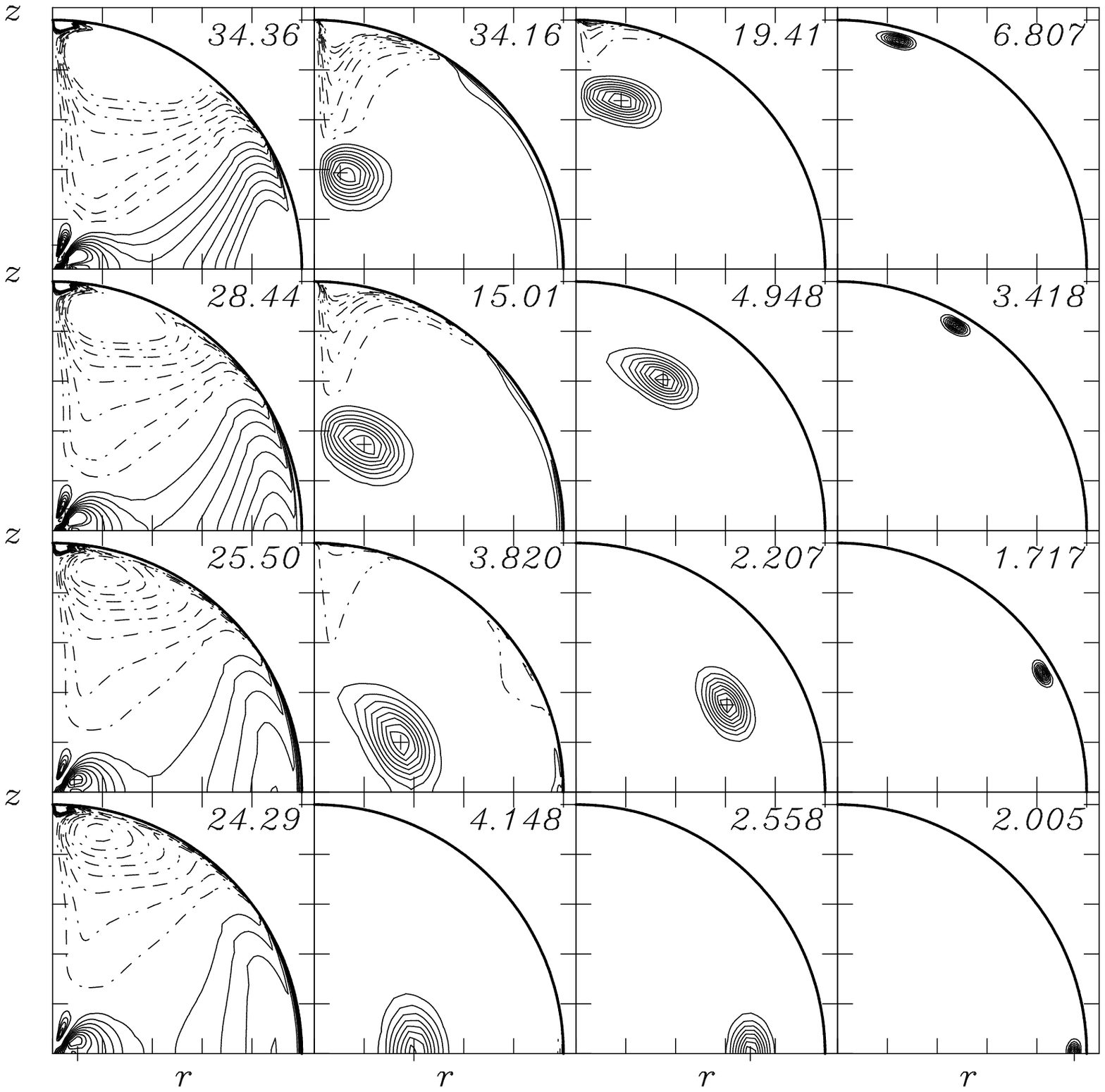}
\caption{{\bf Figure \Taktableau} Unimodular averaging kernels at radii
$r_0 = 0.1, 0.4, 0.7, 0.95$ and co-latitudes $\theta_0 = 15\dgr, 30\dgr,
60\dgr, 90\dgr$, without also localizing the $F_2 G_2$-terms.
Target locations are indicated by crosses; standard errors are as quoted.
Trade-off parameter values are the same as for Fig.~3}
\endfigure\nfig

It should be noted that the GONG Hare and Hounds dataset contains some
low frequency p  and f modes for which the assumptions about the
location of the lower turing point made in the previous section are
inaccurate or inappropriate. Specifically, for f modes
with $l< 13$ and and also for the $l=1, n=1$ p mode, $\zeta^{nl}$ is
computed using \Gfunrat) and \Delteqs) with $r_0$ set to $0.1$ instead
of the turning point radius $r_t$.

The resolution and localization achieved in the inversions can be seen
by inspecting the averaging kernels $\CK(r_0,\theta_0;r,\theta)$,
defined in equation \Avkern). In Fig. {\resolution} we illustrate averaging
kernels at $r_0 = 0.7 R$, $\theta_0 = 45\dgr $, for different choices of
target widths $\Delta_r$ and $\Delta_\theta$. As can be judged from the
crosses superimposed on the kernels, the constructed averaging kernels match
closely the specified target forms. The centre kernel has a radial width
$\Delta_r$ of $0.0648$ (fractional) solar radii and an angular width
$\Delta_\theta$ of $7.75\dgr$. This yields a standard error in the inversion of
$3.3\,$nHz.
Comparing this level of uncertainty
with the range of the rotation rate observed
at the Sun's photosphere (roughly $320\,$nHz -- $460\,$nHz), or the
range (approximately $300\,$nHz -- $1500\,$nHz)
of the rotation rate present in our artificial
example (see Fig. 6a), it is evident that this is an acceptably small error
for many purposes. The resolution could
be squeezed further in either radius or latitude, but at the expense of
increasing the standard error. Conversely, the error could be reduced somewhat
by degrading the radial or latitudinal resolution.

A more complete picture of what may be achieved globally is provided by
Fig. {\ourtableau}, which shows averaging kernels at various target radii
and latitudes. These were all obtained with the same values of the
scaling parameters $\alpha_r = 8.0$, $\alpha_\theta = 1.5$, as well as
the same value of the error trade-off parameter $\mu_0 = 0.1$.
As can be seen, the resolution is best in the outer part of the Sun, as
expected from the asymptotic scalings \Delteqs), because
modes with high $l$ are sensitive to the rotation in these layers. The
standard errors are also smallest in this region. The averaging kernels are
very cleanly localized near the intended target location, for most locations in
the outer half of the Sun. Only near the pole does the kernel exhibit
noticeable nonlocal structure -- only relatively low-$m$ modes sample the
region close to the pole, so it is harder to construct a kernel there. The
standard error is also much higher for the near-polar inversion, for the same
reason.

Even for target locations as deep as $0.4 R$, the equatorial kernels are
very well localized, and the error is only $4\,$nHz. The resolution could of
course be improved further if the error was allowed to increase but, as
illustrated in Fig. {\resolution}, one could not expect to improve the
resolution substantially without at least doubling the standard error.
The polar kernel at this radius shows considerable nonlocal structure, though
it still has a main peak close to the intended location. Thus it is possible
to localize kernels at different latitudes and thus to achieve latitudinal
resolution at this depth, albeit with a considerable uncertainty $\sigma$.
This is not the case at $r_0 = 0.1 R$, our deepest
target radius. Here all the kernels are similar and are localized at low
latitudes. Nonetheless, apart from an obvious positive region in the convection
zone, and a thin negative region near the surface at high latitudes, these
kernels are reaonably well-localized. However, as discussed below, they
depend heavily on the low-degree f modes in the Hare and Hounds dataset,
which have so far not been observed.

To illustrate the importance of the $F_2^{nl} G_2^{lm}$, we have recomputed the
coefficients used to construct the kernels in Fig. {\ourtableau}, setting
$\beta_l = 0$ and $\zeta^{nl} = 0$ for all $n$ and $l$. The values of all other
parameters are unchanged.
The results should
still be superior to a basic $\Rreal^1 \otimes \Rreal^1$ inversion, as
described at the beginning of Section 2, because we are choosing the
coefficients $\clm$ as a function of $r_0$ as well as of $\theta_0$. As
expected, we see in fig. \Taktableau{} that neglecting the
$F_2^{nl} G_2^{lm}$ contribution to the
mode kernels has little effect in the outer part of the Sun, but
degrades the kernels at $r_0 = 0.4 R$ at all but the equatorial target. At the
deepest target locations, the kernels constructed without taking
$F_2^{nl} G_2^{lm}$ into account are very poor indeed. This is precisely as
one would expect, because for such a deep target location the inversion
needs to use low-degree (and, if they are available, low-frequency) modes, for
which it is a poor approximation to neglect $F_2^{nl} G_2^{nl}$.

\fignam{\Coeffs}{Coeffs}
\beginfigure{5}
\epsfxsize=8.5cm
\epsfbox{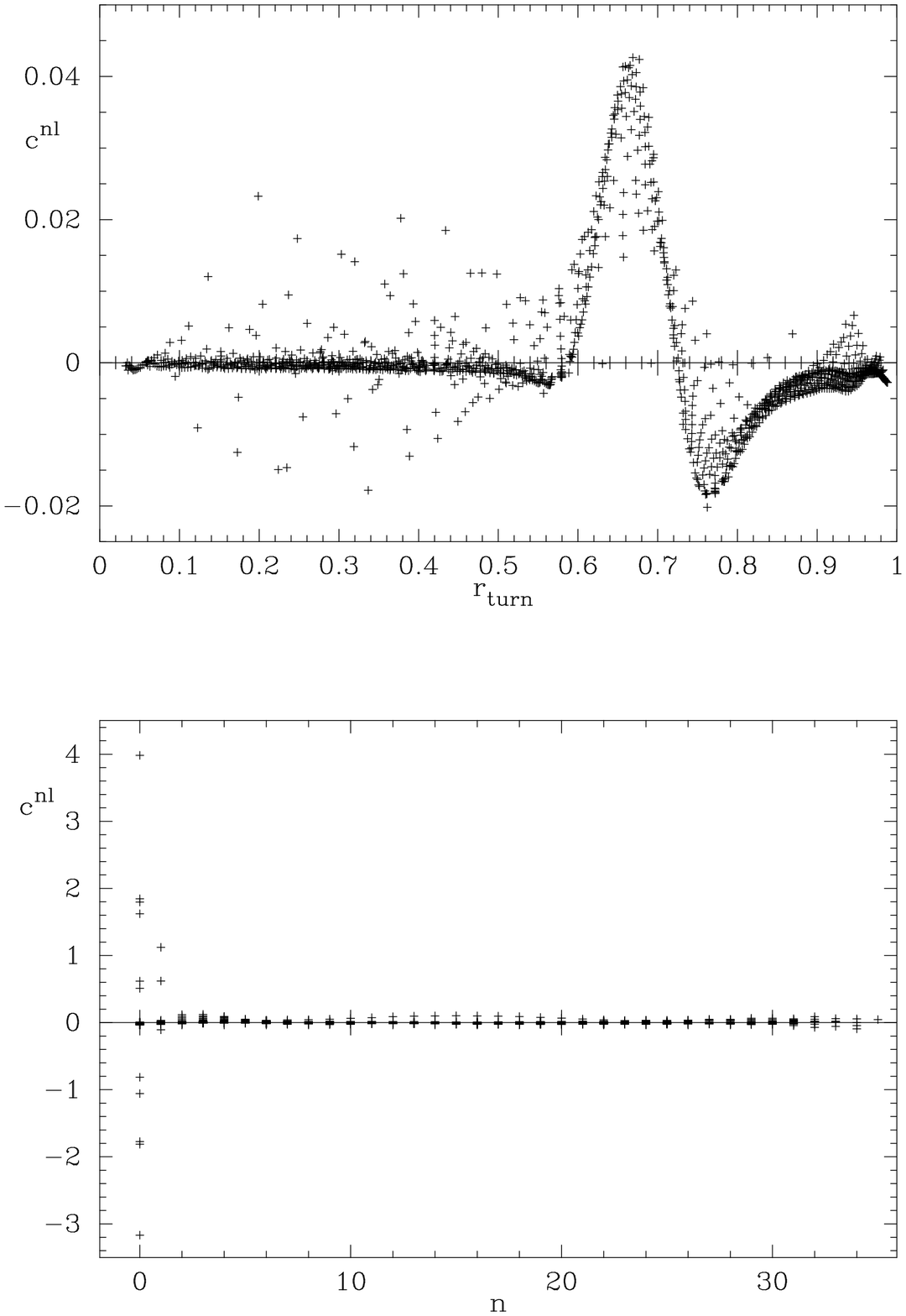}
\caption{{\bf Figure \Coeffs} a. The coefficients $c^{nl}$ for radius
$r_0 = 0.7, \theta_0 = 45\dgr$ as a function of turning point. b. The
coefficients $c^{nl}$ for $r_0 = 0.1, \theta_0 = 90\dgr$ as a function of
degree $n$}
\endfigure\nfig

This point is further illustrated Fig. {\Coeffs}, where the radial coefficients
$\cnl$ are shown for two target locations, using the same parameters
as for Fig. {\ourtableau}. Panel (a) shows coefficients for a target radius
$r_0 = 0.7 R$: in that case, the coefficients are approximately a
function of lower turning point, as has been seen in 1-D inversions
(Christensen-Dalsgaard {\etal} 1990). At $r_0 = 0.1 R$ the low-degree
f modes ($n=0$) and gravest p modes are given most weight, as can be seen in
panel (b). For such modes it is a poor approximation to neglect the
$F_2^{nl}G_2^{lm}$ contribution to the mode kernels, which is why the
kernels in Figs. {\ourtableau} and {\Taktableau} differ so markedly for
such deep target locations.
It could be argued that the f modes should be excluded from the artificial
data-sets because they have not yet been observed. It is clear that
localizing a kernel as deep as $r_0 = 0.1 R$ would then be more difficult
since it is clear from fig. \Coeffs{}b that it is primarily these that
are used in the localization this deep. At $r_0 = 0.4 R$ the coefficients
for the f modes are already of the same magnitude and smaller as all
the other coefficients. Their omission should therefore not significantly
affect the quality of the kernel nor the error magnification.
It is clear from a comparison of figures {\ourtableau} and {\Taktableau}
at $r_0 = 0.4 R$ that the improvement with the new method proposed here is
still appreciable, because of the improved treatment of the gravest
p-modes.

\fignam{\Rotation}{Rotation}
\beginfigure*{6}
\epsfxsize=16.5cm
\epsfbox{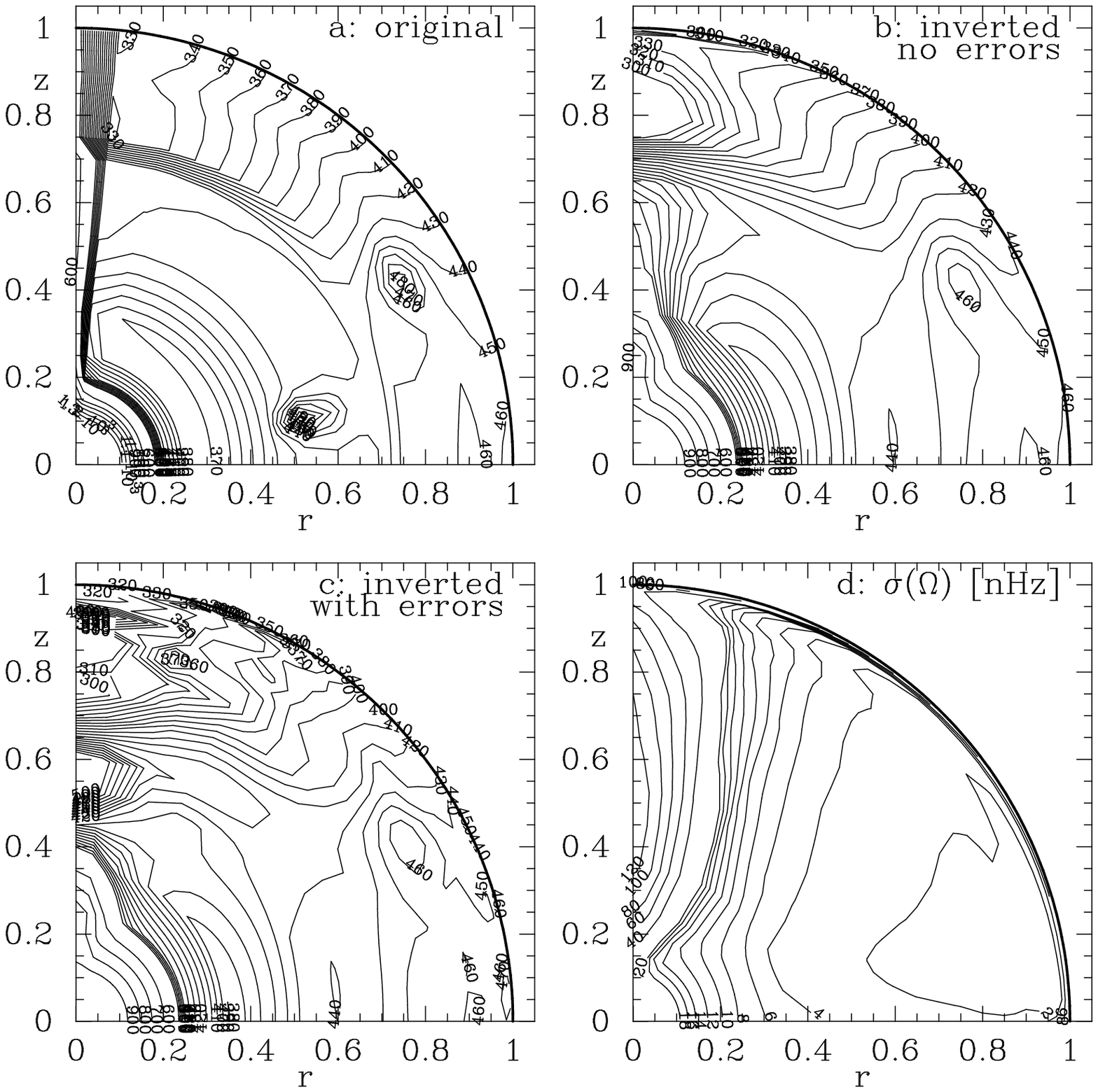}
\caption{{\bf Figure \Rotation} a. The original artificial rotation rate.
b. The reconstruction using error-free data. c. The reconstruction with
noisy data. d. The errors in the reconstructed rotation rate of c.
The parameter vaues used in the reconstructions were
$\alpha_r = 8.0$ $\alpha_\theta = 1.5$ and $\mu_0 = 0.1$.}
\endfigure\nfig

Although the averaging kernels contain complete information about the
resolution of an inversion, it is nonetheless instructive to see how
well our method reconstructs a particular rotation profile. For this
purpose, we have invented a rotation profile, shown in Fig. {\Rotation}a.
Its features include two jets, a fast-rotating core
and polar region, and an enhanced rotation at $r\approx 0.9 R$. Using
the appropriate mode kernels, the
rotation profile was then used to compute artificial splitting data
$D_{nlm}$, which we then inverted using the same parameters as for
Fig. {\ourtableau}. In the inversion, we assumed the error uncertainties
described at the beginning of this section, but we performed two
test cases, one where no errors were actually added to the data and a
second where independent Gaussian errors with the assumed standard
deviations were added as noise to the data. The inversion of the
noise-free data is shown in Fig. {\Rotation}(b). The fast-rotating
core, the abrupt change in rotation at $r\approx 0.7 R$ and the general
trend of the rotation in the outer part of the Sun are all correctly
inferred. The outer jet is rather poorly resolved, and the deeper jet is
not convincingly detected at all. This is consistent with the width
of the main peak of the averaging kernels shown in Fig. {\ourtableau}.
In such a noise-free case, one could of course make much narrower kernels
and hence obtain better resolution, because data errors are not a concern.
However, the main purpose of showing the noise-free inversion is so
that it can be compared with the inversion of noisy data in Fig.
{\Rotation}(c). Except near the pole, the corresponding contours in the
two panels can very clearly be identified. The relatively small differences
are due to the propagation of data errors, and are consistent with the
standard errors quoted in Fig. {\ourtableau} and with the contour plot
of standard errors in Fig. {\Rotation}(d). It is clear from panel (d) that
the standard errrors increase substantially in the near-polar region,
which accounts for the greater distortion of the noisy inversion near
the pole. In the rest of the Sun, however, one can see from
comparing panels (a)-(c) that it might be worthwhile to choose smaller
widths for the averaging kernels, thus improving the resolution,
even at the expense of somewhat greater errors.
\fignam{\Highres}{Highres}
Fig. {\Highres} illustrates that one could make a statistically significant
detection of the deeper jet in this way.

\beginfigure{7}
\epsfxsize=8.5cm
\epsfbox{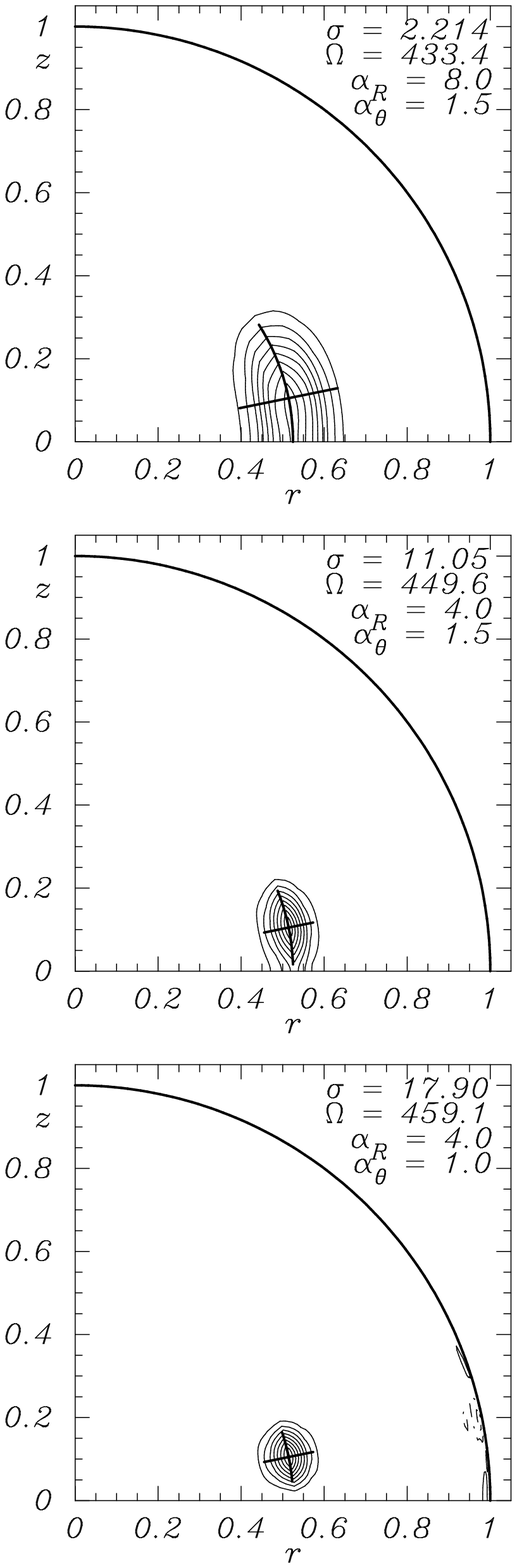}
\caption{{\bf Figure \Highres} The effect of improved resolution on the
error and the deduced rotation rate at the location of the 'jet'}
\endfigure\nfig

\fignam{\Difference}{Difference}
\beginfigure{8}
\epsfxsize=8.5cm
\epsfbox{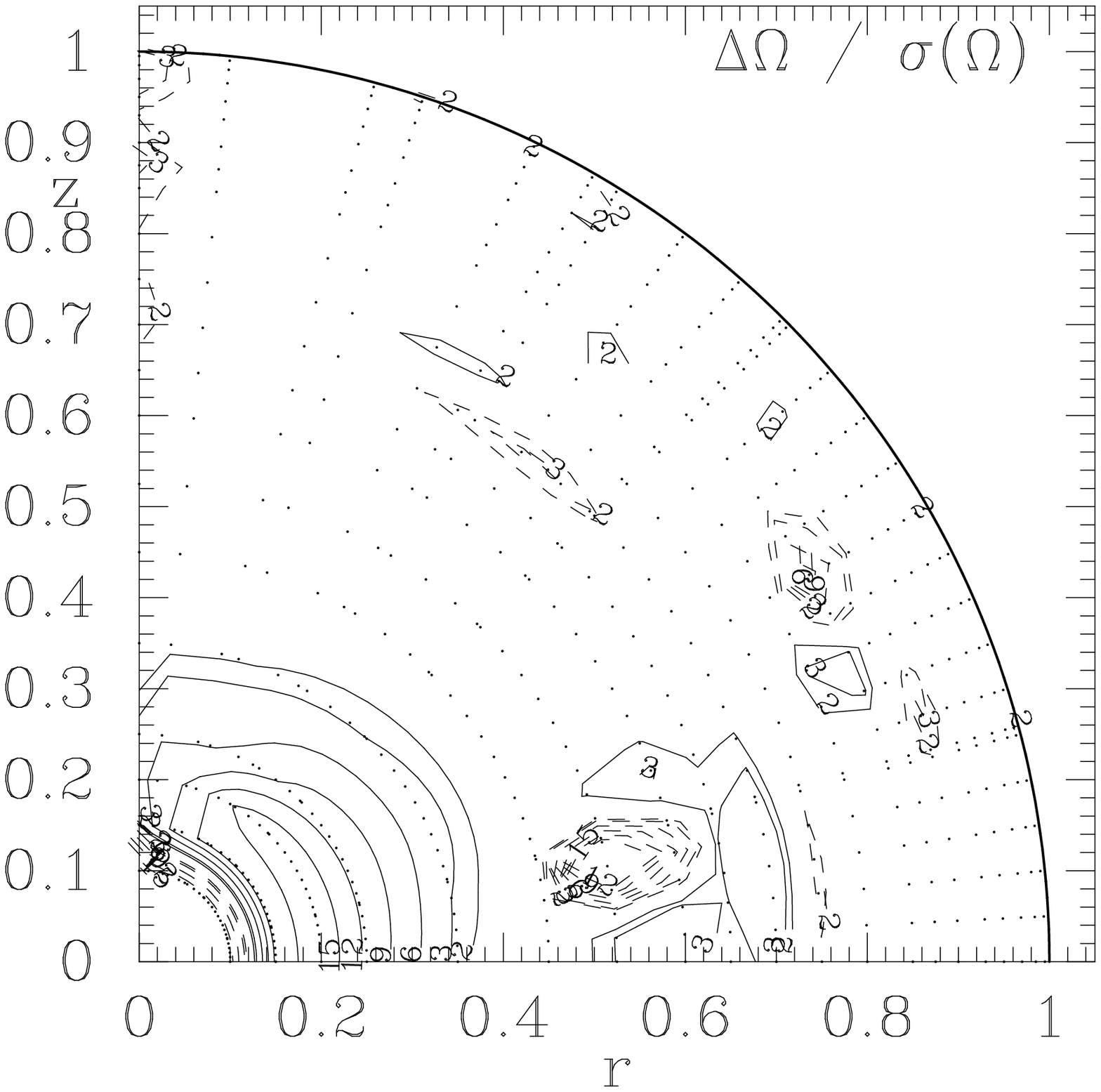}
\caption{{\bf Figure \Difference} The difference between reconstructed
and original, divided at each point by the local $\sigma$. Dots mark
the central locations of all averaging kernels used in the reconstruction
}
\endfigure\nfig

Fig. {\Difference} illustrates the difference between the inferred rotation
from the noisy data and the true rotation profile divided by the
standard error.
In rather few places does the deviation between inferred
and true rotation rise above $2\sigma$. One place is the core,
even though the
inversion reproduces qualitatively the fast rotation. The other most
significant deviations are around the two jets, particularly the deeper
one. In the latter case, the broad averaging kernels have spread the jet
out, so that not only is there a strong negative deviation where the
prograde jet should have been, there is also a positive deviation around
the jet where the inferred rotation is higher than the true rotation.
Also shown in this figure are the locations where we constructed the
inversion solution upon which the contour plots in Fig. {\Rotation}
were based.

\section{Discussion}

We have previously demonstrated (PT1, PT2) that the SOLA method produces
kernels that are of as good a quality as those obtained with the more
traditional OLA formulations and shown that the SOLA method is
computationally more efficient.

Our $\Rreal^1 \otimes \Rreal^1$ inversion differs from the one
proposed by Sekii (1993a,b). He approximated each mode kernel as a product
$F_1^{nl} (r)$ and $G_1^{nl} (\theta)$, where
$F_1^{nl}$, $G_1^{nl}$ are given by equations \Fodef) and \Godef).
The major difference of principle is that Sekii has thus approximated
the kernels, whereas we keep the small term $F_2^{nl} G_2^{nl}$ and
therefore make no approximation (beyond assuming that the rotational
splitting is correctly described by first-order theory). Sekii's
original method would not be recovered completely from ours in the limit
$\beta_l \rightarrow 0$, $\zeta^{nl} \rightarrow 0$, because we do
adjust latitudinal resolution as a function of depth which Sekii (1993b)
already suggested but did not yet implement.

For all but the low degree modes (say $l\lwig 5$), Sekii's approximation
of neglecting $F_2^{nl}$ should be very good. However, to perform
inversions in the deep interior it is highly desirable to include the
lowest degree modes, which penetrate deeply into the star, and this is
apparently a drawback of Sekii's approximation (Sekii 1994). Now
the effect on the inversion in the deep interior of neglecting the
$F_2^{nl}$ term is uncertain a priori,
since even for low degree modes the
$F_2^{nl}$ component is not large compared to $F_1^{nl}$.
However, from figures \ourtableau{} and
\Taktableau{} it can be seen that there is a substantial difference
between kernels at $r_0 = 0.1$ depending on whether or not
the $F_2 G_2$ terms are included.
At this radius, the present method makes substantial use of the
low-degree $n = 0$ modes (Fig \Coeffs{}). It could be argued that
data for such modes are not available at present. But if they are observed
by {\eg} GONG, they will be enormously helpful for probing the core region
and it is highly desirable that our inversion methods make
accurate use of them.

Apart from retaining the $F_2 G_2$ term, and taking this into account in the
latitudinal and radial localizations using SOLA, the present approach is very
much in the same spirit as Sekii's original
$\Rreal^1 \otimes \Rreal^1$ method (Sekii 1993a,b). Subsequently, Sekii has
made a modification to the technique, to make a 2-D integral in the radial
localization (Sekii 1993b, 1994). Such a modification might be included in our
approach also.

Any multiple of $F_2^{nl} G_1^{lm}$ could be added into the
first term on the right-hand side of equation \Separable) and maintain the
form of the right-hand side, provided the same multiple was subtracted from
the second term. In particular we could add $\zeta^{nl} F_2^{nl} G_1^{lm}$
to the first term and subtract it from the second. This is equivalent to
replacing $F_1^{nl}$ with $F_1^{nl}+\zeta^{nl} F_2^{nl}$ and replacing
$G_2^{lm}$ with $G_2^{lm} - \zeta^{nl} G_1^{nl}$. All such operations can be
seen to be linear transformations of the vectors of functions ${\bf F}^{nl}$
and ${\bf G}^{lm}$. In the appendix it is shown that this always leads to
the same inversion procedure provided that the two latitudinal target
functions are equal up to a multiplicative constant. One implication
of this is
that our modified $\Rreal^1 \otimes \Rreal^1$ inversion cannot
be obtained from a linear transformation of Sekii's method, since such
transformations never lead to a radial inversion using just $F_1^{nl}$
as Sekii's method does.

The advantage of both ours and Sekii's $\Rreal^1 \otimes \Rreal^1$ technique
for SOLA type inversions is that it is never necessary to invert a matrix
with a dimension of the order of the total number of frequency splittings
(i.e. $\sim 70 000$). The largest matrix to invert for construction of the
latitudinal averaging kernels is $l_{max}$ which is $250$ for the mode
set under consideration. For the radial averaging kernels it is in
principle the number of different $n,l$-combinations which is 1380 for the
Hare and Hounds mode set. However the number of modes for the radial
inversions can be reduced by projection onto a suitable basis after
performing an SVD reduction (cf. PT1; Christensen-Dalsgaard \& Thompson 1993)
to $87$.

Matrix inversion is an ${\cal O}(N^3)$ process and therefore naively
the speed-up of the inversion due to the technique described here
for this mode set would be a factor close to~:
$$\left( {\sum\limits_{l \in {\cal M}} \left[
N_m(l)^3\right] \over 69662^3}\right)^{-1}
\approx 3\times 10^6 \,,\eqno\neqn $$
where $N_m(l)$ is the number of observed modes with distinct $m$ values
for each $l$. However the integrations in the calculation of the vector
${\bf v}$ in \refeq1) turn out to dominate in the computing time and
this is an ${\cal O}(N)$ process so the overall speed up is more modest.
Taking into account that the full
2-D inversion would require computing 2-D integrals in the determination
of ${\bf v}$ in \refeq1) instead of 1-D integrals the theoretical speed
up is a factor of $\sim 14 \times 500 = 7000$. Here 500 is the
approximate number
of grid points in the radial direction which one would also have to
integrate over in computing a complete 2-D integral (to resolve the
mode kernels adequately), and which can now be
omitted in all 1-D latitudinal inversions. Since there is an integration
for each mode there is an extra factor 14 from the ratio of the
number of modes used in the 2-D and $\Rreal^1 \otimes \Rreal^1$ inversion~:
$$\left({\sum\limits_{l \in {\cal M}} N_m(l) \over 69662 }\right)^{-1}
\approx 14$$
This estimate of the speed-up was not experimentally tested.

Finally it should be noted that a very preliminary version of this
method was used in the GONG Hare and Hounds exercise (Gough \& Toomre,
1993). The results there were poorer than those presented here,
since the weighting factors
$\zeta^{nl}$ and the scheme for obtaining the optimal resolution
at all radii had not been satisfactorily developed at that time.
The quality of the results is sensitive to these choices.

\section*{Acknowledgments}
Much of the work undertaken in this paper was performed while FPP was
supported by grant GR/H33596 from the UK Science \& Engineering Research
Council.

\section*{Appendix A: invariance of kernels under linear transformations}

The 2D $\Rreal^1\otimes\Rreal^1$ splitting of the kernels of equation
\Separable) is invariant under any linear transformation.
The transformations of the vector of functions ${\bf F}$ is written as~:
$$
\left(\matrix{ \Upctil F_1 \cr \Upctil F_2 \cr}\right)\ =\
\left(\matrix{ o & p \cr q & r \cr}\right)
\left(\matrix{ F_1 \cr F_2 \cr}\right)\ ,
\eqno\apneqn
$$
and the transformation of the vector of functions ${\bf G}$ as~:
$$
\left(\matrix{ \Upctil G_1 \cr \Upctil G_2 \cr}\right)\ =\
\left(\matrix{ a & b \cr c & d \cr}\right)
\left(\matrix{ G_1 \cr G_2 \cr}\right)\ ,
\eqno\apneqn
$$
which covers all the transformations considered in this paper.
The inner product of the ${\bf F}$ and ${\bf G}$ vector should not
be affected by the transformation, therefore
$$
{\bf \Upctil F} \cdot {\bf \Upctil G}\ =\
\left(\matrix{ F_1 \cr F_2 \cr}\right)^T
\left(\matrix{ o & q \cr p & r \cr}\right)
\left(\matrix{ a & b \cr c & d \cr}\right)
\left(\matrix{ G_1 \cr G_2 \cr}\right)
\eqno\apneqn
$$
must be equal to~:
$$
\left(\matrix{ F_1 \cr F_2 \cr}\right)^T \left(\matrix{ G_1 \cr G_2
\cr}\right)\ .
\eqno\apneqn
$$
As a consequence of this it is clear that the two matrices must be each
others inverse~:
$$
\left(\matrix{ o & q \cr p & r \cr}\right)\ =\
{1\over (ad-bc)} \left(\matrix{ d & -b \cr -c & a \cr}\right)
\eqno\apneqn
$$
The single kernel used for the radial inversions is $F_R$~:
$$
F_R\ \equiv\ F_1 + \zeta F_2
\eqno\apneqn
$$
Here the constant $\zeta$ is the ratio of the values of the kernels
$\GG$ evaluated at the target location $u=u_0$. In the formulation used
in the main text
$$
\zeta = - \left( {\nu_{\rm min} \over
\nu} \right)^2
\eqno\apneqn
$$
In the transformed set
$$
\Loctil\zeta = {(c + \zeta d ) \over (a + \zeta b)}
\eqno\apneqn
$$
In this case the new radial kernel ${\Upctil F}_R$ becomes
$$
\eqalign{
(a + \zeta b){\Upctil F}_R\ &=\ {1\over (ad-bc)}\left[ (a + \zeta b)
(d F_1 - c F_2) \right. \cr
 &\hskip 1.5truecm \left. + (c + \zeta d) ( -b F_1 + a F_2)\right]\cr
\ &=\ F_1 + \zeta F_2 \cr}
\eqno\apneqn
$$
The factor $(a + \zeta b)$ on the left-hand side drops out again because
of the subsequent normalization of the kernels. The radial part of the
inversion therefore is not affected by the linear transformation.
The latitudinal part of the inversion does appear different but if the
two components of the target function are equal up to a constant factor of
multiplication for the original set $(G_1, G_2)$, then the transformation
will preserve that and therefore also the inversion. \par\noindent
The original method as described in the paper by Sekii (1993b)
which uses the radial kernel $F_R = F_1$ is not obtainable
from an invertible linear transformation of the kernels.
\end
\bye